\documentclass[aps,pre,twocolumn,superscriptaddress]{revtex4-2}

\usepackage[usenames,dvipsnames,x11names]{xcolor}
\usepackage[normalem]{ulem}
\usepackage{graphicx}
\usepackage{listings}
\usepackage{stmaryrd}
\usepackage{amssymb}
\usepackage{amsmath}
\usepackage{gensymb}
\usepackage{float}
\usepackage{bbold}
\usepackage{bm}
\usepackage[linkcolor=blue,citecolor=blue,urlcolor=blue,colorlinks=true]{hyperref}

\newcommand{\Tr}{{\rm Tr}}

\begin{document}
\title{Spatial Control of Compressible Active Fluids}

\date{\today}

 \author{Yu-Chuan Cheng$^{1}$, Dimitris Krommydas$^{1}$, Seungwoo Shin$^{1}$, Paarth Gulati$^{2,1}$, Michael F. Hagan$^{3}$, \\ Aparna Baskaran$^{3}$, Cristina M. Marchetti$^{1,4}$, Zvonimir Dogic$^{1,4}$ \\ \normalsize $^{1}$\textit{Department of Physics, University of California, Santa Barbara, Santa Barbara, California 93106, USA}\\ \normalsize $^{2}$\textit{Department of Physics, Emory University, Atlanta, GA 30322, USA} \\ \normalsize $^{3}$\textit{Department of Physics, Brandeis University, Waltham, MA 02453, USA} \\ \normalsize $^{4}$\textit{Interdisciplinary Quantitative Biosciences, University of California, Santa Barbara, CA 93106, USA}\\   }

\begin{abstract}
We study the response of compressible microtubule-based active fluids to sinusoidally varying activity patterns, using experiments and theory. Microtubules accumulate in low-activity regions and are depleted from high activity ones, with the density difference controlled by the wavelength of the activity pattern. To explain these observations the continuum model needs to account for both active stresses that are proportional to nematic order tensor and the isotropic active pressure. At short wavelengths, active vortices spanning adjacent high and low activity regions suppress density modulation. At larger wavelengths, vortical flows confined to the low or high activity region allow for the emergence of the density contrast. Our work shows that a minimal theoretical model of active compressible fluids accurately reproduces experiments, while elucidating the fundamental constraints on control protocols that seek to tame their chaotic dynamics.
\end{abstract}

\maketitle

\section{Introduction}
Cytoskeletal active fluids are assembled from energy-consuming motor proteins and associated filamentous structures, which collectively generate mesoscale active stresses. In conventional realizations of active fluids, active stresses are generated uniformly throughout the sample and drive self-sustained turbulent-like flows~\cite{aditi2002hydrodynamic,saintillan2007orientational,sanchez2012spontaneous,kumar2018tunable}. Geometrical confinement transforms bulk chaotic states into circular coherent flows~\cite{wioland2013confinement,wu2017transition}, but does not allow reconfiguration between different dynamical states. The same energy-consuming proteins that form engineered active fluids drive transport in living cells. Living matter, however, is able to control active stresses in space and time to achieve functions such as morphogenesis, cell motility, and cell division~\cite{brugues2014physical,mayer2010anisotropies,mitchell2022visceral,Guglielmi2015}.  This observation has inspired the creation of light-controllable active matter, where the local magnitude of the mesoscale active stresses can be tuned with light~\cite{Ross2019Nature,peng2021imaging,yang2025dynamic,LinneaPNASnexus}. The imposed light pattern can be dynamically programmed, creating the potential for applications where active flows can be designed in space and time. Achieving this potential requires a quantitative understanding of how extensile fluids respond to spatially varying activity patterns. 

A significant challenge is a disconnect between theoretical models and experiments. On the one hand, numerous realizations of extensile active fluids, ranging from microtubule-kinesin mixtures to bacterial suspensions, are two-component systems: the active agents form a compressible medium within an incompressible passive solvent~\cite{wensink2012meso,sanchez2012spontaneous,zhou2014living,kumar2018tunable,peng2021imaging}. Most theoretical models, on the other hand, adopt a one-fluid description in which the velocities of the active agents and the background passive fluid are described by the same incompressible coarse-grained field which ensures uniform density of active agents~\cite{marchetti2013hydrodynamics}. However, spatially patterned active matter couples activity and density. Motile units from activated regions rapidly migrate into an adjacent low-activity region. In comparison, immotile units from low-activity regions hardly ever cross into high-activity regions. Such imbalances generate concentration gradients~\cite{arlt2018painting,frangipane2018dynamic,LinneaPNASnexus}, which are incompatible with theoretical models that treat the active units and background fluid with the same velocity. Recent work has begun to address this gap~\cite{yang2025dynamic}, but has not included the role of liquid crystalline degrees of freedom and associated deviatoric active stresses that are known to drive active turbulence. 

We study an optically responsive, compressible, microtubule-based extensile active fluid. We examine the response to sinusoidally modulated activity, which reveals the emergence of density patterns and spatial modulation of autonomous flows. The magnitude of the microtubule density contrast between dark (passive) and illuminated (active) regions is controlled by the wavelength of the applied sinusoidal light pattern, while the shape of the density profile is controlled by the sharpness of the activity gradients. We show that a minimal theoretical model explains these observations ~\cite{krommydas2026compressible}, while revealing the complementary roles of the deviatoric and isotropic components of the active stress. The isotropic active stress provides an active pressure that pushes microtubules out of the high-activity regions and controls the density modulations. The deviatoric active stress controls the active flows, which suppress density modulations when the wavelength of the light modulation becomes shorter than the correlation length of vortical flows. Our work provides a tractable theoretical model that captures the spatially patterned dynamics of compressible active fluids.  

Previous efforts have demonstrated external open-loop control of contractile networks with optically engineered motor clusters or light-activated motor inhibitors~\cite{schuppler2016boundaries, linsmeier2016disordered,Ross2019Nature}. These systems can be light-activated to undergo a one-time deterministic transition from an expanded to a contracted state, but the transition is not reversible. In contrast, extensile fluids exhibit a steady state chaotic dynamics. Controlling such chaotic dynamics provides a unique opportunity to engineer dynamical states that can be reversibly tuned by modulating the light pattern and for controlling the dynamics of motile topological defects~\cite{norton2020optimal,zarei2023light,shankar2024design,nishiyama2025closed}. Our findings provide a foundation for controlling the spatiotemporally chaotic dynamics of compressible active fluids using external feedback, while also elucidating fundamental constraints on such protocols. 

\section{Optical Control Of Activity}
\begin{figure}[htbp!]
\centering
\includegraphics[width=0.49\textwidth]{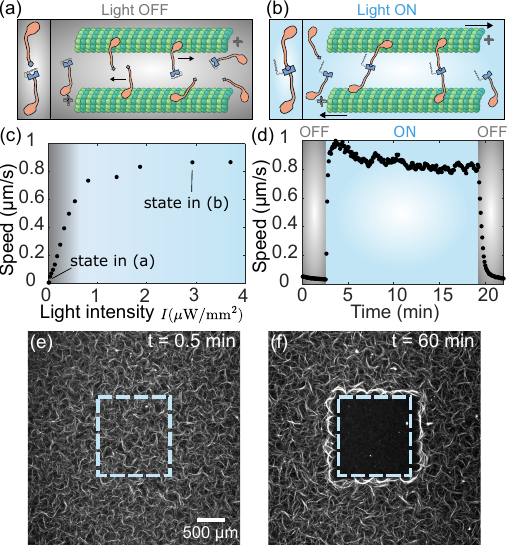}
\caption{\textbf{Optically-responsive MT-based active isotropic fluid.}
(a,b) Single-headed kinesin motor domains are fused to optically-responsive iLID and micro domains. In the dark, the motor domains remain unclustered, generating little interfilament sliding. Upon blue-light illumination, motors cluster, cross-link adjacent microtubules, to drive interfilament sliding. (c) Spatially averaged speed of autonomous flows as a function of light intensity. (d) Switching blue light on and off reveals the temporal response of the autonomous flows. (e) Microtubule-based active fluid, immediately after the activity is applied within the boxed region. The reminder of the sample is passive. (f) After one hour, microtubules evacuated the illuminated region, due to the low compressibility of the extensile active fluid.}
\label{fig:1}
\end{figure}

Conventional microtubule-based active fluids are powered by kinesin motor fragments that are fused into clusters by biotin-streptavidin bond~\cite{nedelec1997self}. In presence of depletion interactions, these clusters simultaneously bind adjacent microtubules to drive interfilament sliding. In turn, such microscopic dynamics cascades over several hierarchical levels to generate large-scale chaotic flows~\cite{sanchez2012spontaneous}. In this work, we use motors whose clustering is controlled by blue light~\cite{Ross2019Nature,LinneaPNASnexus}. Upon illumination, two complementary kinesin fragments dimerize. Only dimerized motors bind adjacent microtubules to drive their sliding [Fig.~\ref{fig:1}(a,b)]. Light controls the fraction of clustered motors, which in turn determines the magnitude of the active stresses and active flows. Consequently, the mean speed of the autonomous flows increases with light intensity, approaching a plateau at high intensity [Fig.~\ref{fig:1}(c)]. The light-induced activation is reversible: alternating light on and off switches between an active flowing state and a nearly quiescent state, with characteristic activation and deactivation times of approximately 50~s [Fig.~\ref{fig:1}(d)]. We have optimized the concentration of opto-kinesin clusters so that the dark state has negligible flow speeds.

We study active fluids in $100~\mu\mathrm{m}$-tall chambers with lateral ($x-y$)  dimensions $5.0 \times 20.0 ~\mbox{mm}$. The concentration of filamentous microtubules is approximately 0.3 mg/ml. Suspensions at such low concentration  are isotropic and  highly compressible. Activity generates local nematic order, which in turn generates active flows ~\cite{santhosh2020activity,srivastava2016negative}. Spatial activity patterns were imposed by projecting blue activation light using a digital micro-mirror device through diascopic illumination. The active fluid dynamics of red-labeled microtubules was quantified with an epi-fluorescence method~\cite{zarei2023light}. Our setup allowed us to continuously excite the entire centimeter sized sample with blue light while observing any sample sub-region region with red light and to raster scan the entire sample. The depth of focus of the projected illumination is  $\approx 1~\mathrm{mm}$, much larger than the chamber height, so the activation profile is nearly uniform across the sample thickness.  

\section{Wavelength-dependent density contrast}
\begin{figure}[htbp!]
\centering  
\includegraphics[width=0.49\textwidth]{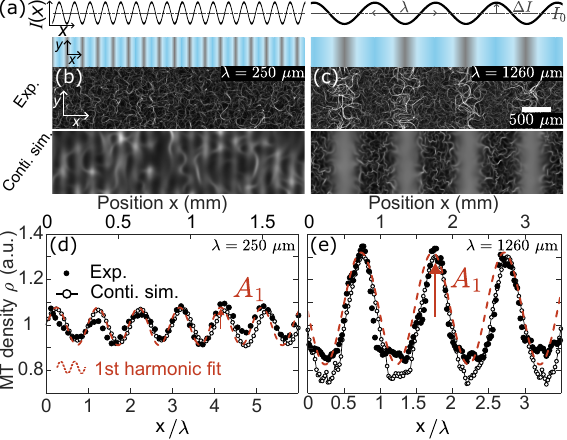}
 \caption{\label{fig:2} \textbf{Activity wavelength controls density contrast.} \ (a) Imposed sinusoidal pattern with wavelength $\lambda$, mean intensity $I_0$, and modulation amplitude $\Delta I$. Two dimensional sinusoidal light pattern imposed on the sample plane, with uniform intensity along y and periodic modulation along x. (b,c) Experimental MT density modulation after $1.5~\mathrm{h}$ for $\lambda=250~\mu\mathrm{m}$ and $1260~\mu\mathrm{m}$ is shown in the top row. The bottom rows show corresponding snapshots from continuum simulations, respectively. (d, e) Steady-state $y$-averaged density profile $\rho(x)$. The dashed red curve is the first-harmonic fit with amplitude $A_1$ characterizing density contrast.}
\end{figure}

Illuminating an extensile fluid of uniform density over a millimeter sized area activates local flows (Fig.~\ref{fig:1}(e)). Over a period of one hour the microtubules evacuate the illuminated region (Fig.~\ref{fig:1}(f)). Once the bundles move to the dark region, their activity ceases; they remain in the inactive region  generating a permanent density contrast. To quantify how such activity-density coupling depends on the  length scale of the illumination pattern, we apply a sinusoidal illumination with varying wavelength: $I(x,y)=I_0+\Delta I \ \rm{sin(2\pi x/\lambda)}$. Here, $x$ is the direction of the sinusoidal variation [Fig.~\ref{fig:2}(a)], while the value $I_0$ is chosen as the midpoint of the regime where the flow speed depends linearly on the imposed intensity (Fig.~\ref{fig:1}(c)). 

Upon illumination with a 250 $\mu$m-wavelength pattern, a uniform sample develops a spatially varying microtubule density. Microtubule accumulate in the low intensity regions; the sample exhibits chaotic flows throughout space [Fig.~\ref{fig:2}(b)]. Increasing the wavelength to 1260 $\mu$m generates more pronounced differences between the maximum and minimum microtubule density [Fig.~\ref{fig:2}(c)]. To quantify the density contrast, we measure the $y$-averaged MT density profile, $\rho(x)$ [Fig.~\ref{fig:2}(d,e)].  

\begin{figure}[htbp!] 
\centering
\includegraphics[width=0.49\textwidth]{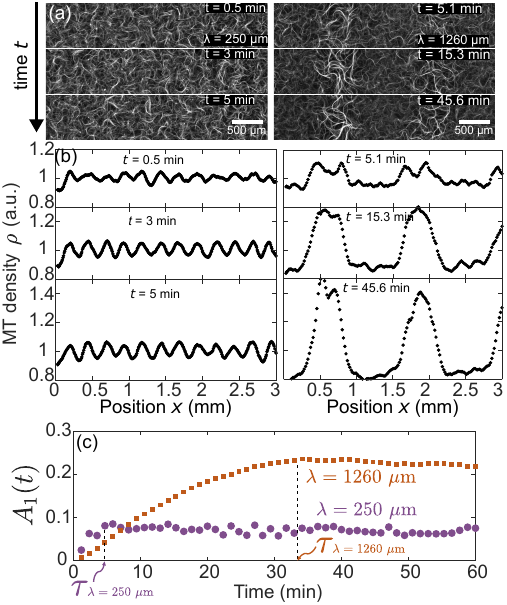}
\caption{\label{fig:3} 
\textbf{Temporal dynamics leading to steady state} (a) Temporal development of the microtubule (MT) density  modulation, for $\lambda=250~\mu\mathrm{m}$ and $1260~\mu\mathrm{m}$. (b) Corresponding $y$-averaged density profiles $\rho(x)$. (c) The evolution of first-harmonic amplitude $A_1(t)$ as a function of time for different imposed wavelengths. Dash lines yield characteristic wavelength dependent time $\tau_\lambda$.}
\end{figure}

The density contrast emerges from a uniform suspension over a characteristic wavelength-dependent time scale [Fig.~\ref{fig:3}a,b]. To quantify this time, we measure the $y$-averaged density profile $\rho(x,t)$ at time $t$; the light patterns is applied at $t=0$ [Fig.~\ref{fig:3}(b)]. We expand $\rho(x,t)$ in a harmonic series and use the first-harmonic amplitude $A_1(t)$ as a measure of the instantaneous density contrast. The time-dependent density contrast follows a saturating exponential growth [Fig.~\ref{fig:3}(c)]. To extract the characteristic time $\tau$ we fit the measured profile to: $A_1(t)=A_1\left(1 - e^{-t/\tau}\right)$ where $A_1$ is the steady-state amplitude and $\tau$ is the characteristic time (Fig. S4).  For 250 $\mu$m wavelength, the characteristic time $\tau$ is a few minutes [Fig.~\ref{fig:3}c]. Increasing the wavelength to 1260 $\mu$mm increases $\tau$ to $\approx 20$ minutes.

\begin{figure*}[htbp!] 
\centering
\includegraphics[width=0.90\textwidth]{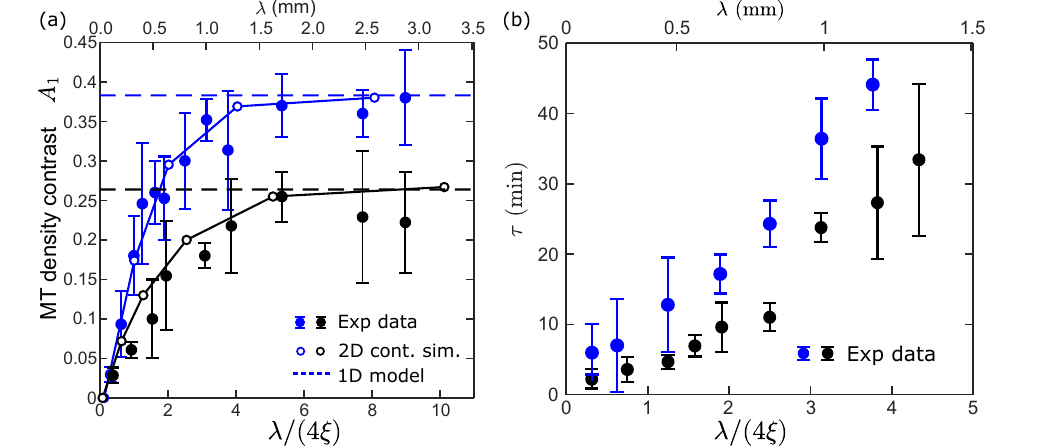}
 \caption{\textbf{Wavelength-dependent density contrast and timescale of dynamics.}
(a) Density contrast $A_1$ as a function of $\lambda/(4\xi)$. Black and blue curves correspond to different modulation amplitudes, $\Delta I=0.3$ and $0.5~\mu\mathrm{W}\ /\mathrm{mm}^{2}$, respectively, with fixed mean intensity $I_0=0.5~\mu\mathrm{W}\ /\mathrm{mm}^{2}$. The solid curves are obtained from the full continuum model; high-amplitude data \((\alpha,\alpha_B)=(-7.7,-2.45)\); low-amplitude data \((\alpha,\alpha_B)=(-4.62,-1.47)\). The horizontal lines are predictions of the minimal diffusive model (Eq.~\eqref{eq:diffusion}), with $R=1.17$ and $R=0.70$ for the high- and low-amplitude cases, respectively. (b) Time characteristic time $\tau$ required to reach steady state is plotted as a function of the dimensionless wavelength $\lambda/(4\xi)$. }
\label{fig:a_1}
\end{figure*}

Next, we systematically vary the wavelength to quantify how the steady-state density contrast depends on the wavelength of the activity profile [blue curve - Fig.~\ref{fig:a_1}(a)]. We normalize the imposed wavelength by the intrinsic flow correlation length $\xi$, measured under uniform illumination (Fig. S1). For  patterns of very short wavelengths we observe no contrast. With increasing wavelength the density contrast increases. For large wavelength the density contrast plateaus at $A_1 \approx 0.25$. We repeat the experiments for a larger modulation amplitude $\Delta I$, for which the peak illumination intensity approaches the saturation regime, while the minimum activity corresponds to dark state [black curve - Fig.~\ref{fig:1}(c)]. For larger amplitude $\Delta I$, the dependence of the density contrast on the wavelength exhibits the same functional behavior but saturates at the higher value of $A_1 \approx 0.4$. 

The characteristic timescale to develop steady state $\tau$ increases with increasing wavelength $\lambda$: longer-wavelength patterns take more time to develop (Fig.~\ref{fig:a_1}(b)). The characteristic time $\tau$, depends linearly on the wavelength $\lambda$ (Fig.~\ref{fig:a_1}(b)), which yields the effective redistribution speed $v_{\mathrm{eff}}\simeq 0.3~\mu\mathrm{m}\,\mathrm{s}^{-1}$. Such speed is comparable to the spatially averaged MT flow speed, $v_{\mathrm{MT}}\simeq 0.8~\mu\mathrm{m}\,\mathrm{s}^{-1}$ of uniformly illuminated sample. Such scaling suggests a mass redistribution over a length set by the imposed pattern. The redistribution of microtubules is governed by the competition between the length scale of the imposed activity pattern and the intrinsic scale of active flows. For sinusoidal illumination, the characteristic redistribution length is $\lambda/4$. Normalizing by the flow correlation length $\xi$, measured under uniform illumination, yields the dimensionless ratio $\lambda/(4\xi)$.

\begin{figure}[htbp!] 
\centering
\includegraphics[width=0.49\textwidth]{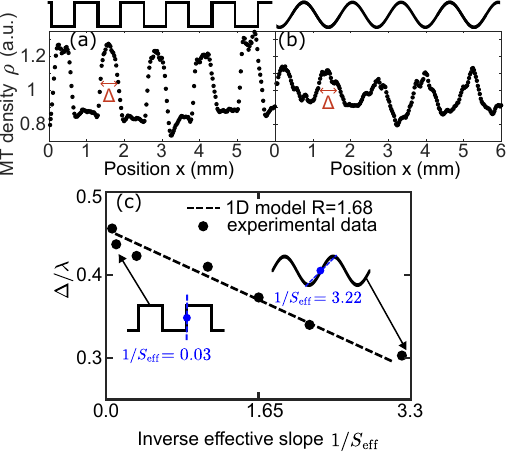}
 \caption{\textbf{Sharpness controls MT density compression.} (a,b) Representative steady-state MT density profiles generated by square-wave-like and sinusoidal activity patterns, respectively. (c) Normalized density-peak width, $\Delta/\lambda$, as a function of inverse effective sharpness, $1/S_{\rm eff}$. Sharper activity interfaces generate broader, more plateau-like density peaks, whereas smoother activity profiles produce narrower, more compressed peaks.}
\label{fig:asymm}
\end{figure}

The steady-state density profiles are asymmetric: the low-density valleys are broad and shallow while the high-density peaks are steep and narrow; such asymmetry becomes more pronounced at larger wavelengths [Fig.~\ref{fig:asymm}(a,b)]. To quantify this asymmetry, we define the peak width $\Delta$ as the distance between the two inflection points of a high-density peak. The inflection points are extracted from a smooth reconstruction of the density profile using the first six spatial Fourier harmonics (Fig. S3). 

To elucidate the features that control asymmetry of the density profile we imposed a periodic activity profile $I(x)=-\,|I|\left[1+\tanh\!\left(S_{\rm{eff}}\,\sin\!\left[2\pi x/\lambda\right]\right)\right] /2$ at fixed wavelength $\lambda=1~\mathrm{mm}$ and fixed intensity amplitude, while varying only the sharpness parameter $S_{\rm{eff}}$ [Fig.~\ref{fig:asymm}(c)]. Small values of $1/S_{\rm eff}$ correspond to sharp, square-wave-like activity patterns , whereas large values of $1/S_{\rm eff}$ correspond to smoother, nearly sinusoidal profiles [Fig.~\ref{fig:asymm}(b)]. Sharper activity patterns produce broader, more symmetric density profiles, with $\Delta/\lambda$ approaching $0.5$, while smoother more sinusoidal activity patterns generate narrower, more compressed and asymmetric peaks [Fig.~\ref{fig:asymm}(c)]. Thus, even at fixed wavelength and activity amplitude, the sharpness of the activity pattern controls the asymmetry and compression of the steady-state MT density profile.  

\begin{figure}[htbp!] 
\centering \includegraphics[width=0.40\textwidth]{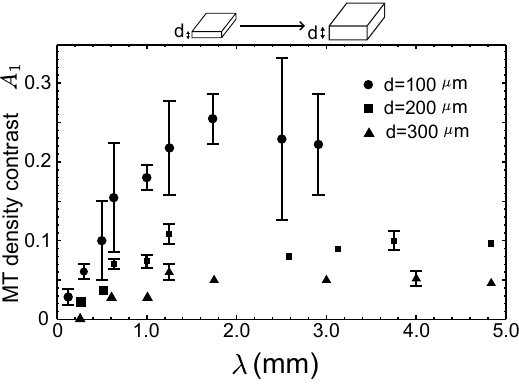} 
\caption{\textbf{Chamber thickness affects density contrast.} Steady-state MT density contrast $A_1$ as a function of the imposed wavelength $\lambda$ for chambers with different thickness. The contrast increases with wavelength and saturates at large $\lambda$. Increasing the chamber thickness systematically reduces the contrast and lowers the plateau value. Data taken with light modulation amplitude, $\Delta I=0.3~\mu\mathrm{W}\ /\mathrm{mm}^{2}$, with mean intensity $I_0=0.5~\mu\mathrm{W}\ /\mathrm{mm}^{2}$.} \label{fig:thickness} 
\end{figure}

We next varied the chamber thickness to test how vertical confinement controls light-induced density contrast. For each thickness, we imposed a sinusoidal activation pattern with wavelength $\lambda$ and measured the steady-state contrast of the $y$-averaged MT density profile. For the 100 $\mu$m thick chambers, the contrast  increases with $\lambda$, and approaches a plateau at long wavelengths [Fig.~\ref{fig:thickness}]. Increasing the chamber thickness to 200 $\mu$m weakens the density modulation, producing a lower contrast and a smaller plateau value, while for 300 $\mu$m the contrast almost disappears. At the same time, the velocity correlation length increases with chamber thickness (Fig. S1). Thus, chamber thickness provides an additional parameter that controls the amplitude of optical density patterning: weaker confinement reduces the efficiency with which the imposed activity pattern is converted into a steady-state density contrast. 

\section
{Continuum model of extensile fluids with patterned activity}
A complete continuum theory of active microtubules in a passive solvent requires a three-dimensional theoretical model that account for two-fluids. We take advantage of the quasi-$2D$ experimental geometry to construct an effective one-fluid hydrodynamic theory~\cite{krommydas2026controlling}. In the underlying two-fluid description, the active microtubule phase and passive solvent have distinct in-plane velocities, while incompressibility constrains only their combined volume flux. Relative motion between microtubules and solvent therefore allows the microtubule phase to compress while the total suspension remains incompressible. In the dilute, weak-interphase-coupling limit, the solvent can be eliminated to obtain the local equations used below in terms of the microtubule concentration $\phi=\rho/\rho_0$, with $\rho$ and $\rho_0$ the local and mean densities, the microtubule velocity $\mathbf{v}$ and the nematic tensor $Q_{ij}=S(n_in_j-\delta_{ij}/2)$. The scalar order parameter $S$ and the director $\mathbf{n}$ capture the magnitude and orientation of local alignment. The equations are given by
\begin{align}
&\partial_t \phi +  \boldsymbol\nabla \cdot \mathbf{v} \phi = D \nabla^2\phi \;,
\label{eq:mass}\\
&\eta \nabla^2\mathbf{v} +\boldsymbol{\nabla} \cdot \boldsymbol{\sigma}^a - \boldsymbol{\nabla}(p+p_a) = \zeta \phi  \mathbf{v}\;,
\label{eq:momentum}\\
&\partial_t \mathbf{Q}  + \mathbf{v} \cdot\boldsymbol\nabla \mathbf{Q}= \lambda_N   \mathbf{A} - [\boldsymbol{\omega}, \mathbf{Q}]+ \frac{1}{\gamma}\mathbf{H}\;.
\label{eq:nematic}
\end{align}
The first equation describes conservation of microtubules, with $D$ their diffusivity relative to the ambient fluid. The velocity field satisfies the Stokes equation, Eq.~\eqref{eq:momentum}, where $\eta$ is the shear viscosity and $\zeta$ is an effective friction per unit area. The deviatoric active stress is written as
\begin{align}
\boldsymbol{\sigma}^a = \alpha(\mathbf{r})\phi\mathbf{Q}\;,~~~~~~\alpha(\mathbf{r})=-\alpha_0f(\mathbf{r})\;,
\label{eq:sigma-a}
\end{align}
where the extensile activity has amplitude $\alpha_0>0$ and is spatially modulated by the function  $f(\mathbf{r})$. The pressure has both passive and active contributions $p$ and $p_a$. The effective microtubule-phase pressure has both passive and active contributions $p$ and $p_a$. The passive pressure is taken to be ideal-gas-like, $p =  \phi /\kappa $, with $\kappa$  the compressibility. This is the osmotic pressure of the microtubules and should be distinguished from the mixture pressure that enforces incompressibility of the total suspension in the underlying two-fluid description. The active pressure is $p_a=-\alpha_B(\mathbf{r})\phi$, with $\alpha_B(\mathbf{r})=-\alpha_B^0 f(\mathbf{r})$ and $\alpha_B^0>0$. To reduce the number of parameters, we take  $\alpha_B^0=\frac{1}{3}\alpha_0$. Finally, the dynamics of the nematic order parameter is governed by the interplay between flow and elasticity, with $A_{ij} =(\partial_i v_j + \partial_j v_i - \delta_{ij} \partial_k v_k)/2$ the traceless and symmetrized strain rate, $\omega_{ij} = (\partial_i v_j -\partial_j v_i)/2$  the vorticity, $\lambda_N$ the flow alignment parameter and $\gamma$ the rotational viscosity. The molecular field $\mathbf{H}=-\delta F_{\mathrm{LdG}}/\delta\mathbf{Q}$ is obtained from the usual Landau--de Gennes free energy, $F_{\mathrm{LdG}} = \int d\mathbf{r} f_{\mathrm{LdG}}$, with $f_{\mathrm{LdG}} = r(\phi)\Tr[\mathbf{Q}^2]/2 + u(\phi)\Tr[\mathbf{Q}^2]^2/4 + K(\nabla \mathbf{Q})^2/2$, where $r(\phi)=r_0\left(1-\phi/\phi_{IN}\right)$  and
$u(\phi)=u_0\phi/\phi_{IN}$. The phenomenological constants $r_0,u_0>0$ set the shape of the Landau potential, while $K>0$ is the elastic stiffness penalizing gradients in nematic order. As in the experiments, we focus on a dilute, homogeneous, isotropic initial state, corresponding to $\phi_0<\phi_{IN}$. Our model departs from one-fluid descriptions of nematic hydrodynamics by allowing for spatial variations of the microtubule concentration, generated by pressure and activity gradients~\cite{marchetti2013hydrodynamics,marenduzzo2007steady,giomi2014defect,giomi2015geometry,ruske2022activity}. Key parameters in the model are the compressibility $\kappa$ (or its inverse, the bulk modulus $B=\kappa^{-1}$), the activity strength $\alpha_0$, and the length scale that controls the modulation of the activity pattern. The other parameters are kept fixed for different conditions (Methods).

We numerically integrated the hydrodynamic equations using a pseudospectral solver on a two-dimensional grid of size $L\times L$. Lengths are measured in units of the bare liquid-crystal coherence length $\ell_c=\sqrt{K/r_0}$, times in units of the nematic relaxation time $\tau=\gamma/r_0$, and energies in units of the elastic constant $K$, such that the integration $\Delta x=\Delta y=0.5$ and $\Delta t=1\times10^{-4}$. All simulations are run for at least $5\times10^5$ steps, after which observables are reported in the dynamical steady state. 

\subsection{The one-fluid compressible hydrodynamics captures the modulation of active flows.}
\begin{figure}[htbp!]
\centering
\includegraphics[width=0.485\textwidth]{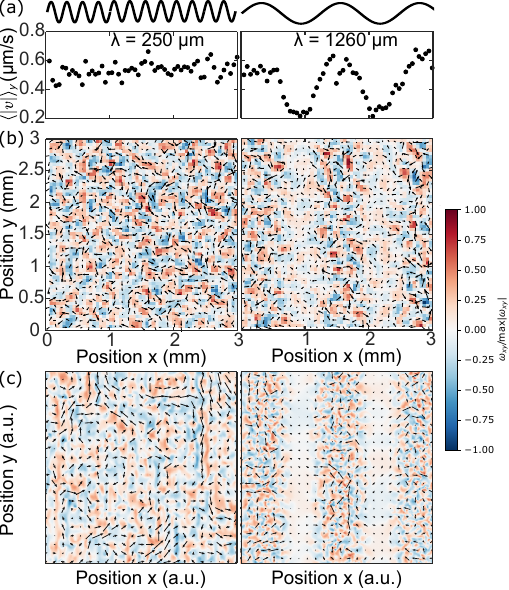}
\caption{\textbf{Spatial organization of the velocity and vorticity fields.}
\textbf{(a)} The modulation of $y$-averaged speed, $\langle |v| \rangle_y(x)$  increases with the wavelength. \textbf{(b,c)} Vorticity maps of active microtubule flows from experiments (b) and simulations (c) under short- and long-wavelength illumination. Arrows indicate the local fluid velocity direction. For $\lambda=250~\mu\mathrm{m}$, vortices span both bright and dark regions, leading to enhanced transport and fluid mixing. In contrast, for $\lambda=1260~\mu\mathrm{m}$, vortical flows are largely confined to the bright regions, whereas the dark regions exhibit weaker flows. For comparison between experiments and simulations, the vorticity in each map is normalized by its maximum absolute value.}
\label{fig:vortex}
\end{figure}

We have solved numerically Eqs.~(\ref{eq:mass}-\ref{eq:nematic}) for a uniform isotropic initial state with a sinusoidal pattern of activity $f(x)=[1+\sin 2\pi x /\lambda]/2$. The theory predicts density accumulation in the low activity regions (Fig.~\ref{fig:2}), while chaotic vortical flows remain finite throughout the system, albeit weaker in the low activity regions (Fig.~\ref{fig:vortex}). Density variations are driven by the extensile active pressure $p_a$. This yields a spatially dependent effective bulk modulus $B_a(x)=B+|\alpha_B(x)|$ that pushes fluid away from the high-activity regions overcoming the lower bulk modulus of the low-activity regions. The deviatoric active stress generates spatio-temporal chaotic flows tuned by the spatial variation of activity.  

Similar to experiments, we have expanded the $y$-- and time-averaged density profile $\phi(x)$ in harmonics to obtain the density contrast $A_1$ as a function of the wavelength $\lambda$ of the activity patterns (Fig.~\ref{fig:a_1}(a)). Predicted  $A_1$ increases linearly at small wavelengths and plateaus at large wavelengths. The crossover wavelength is controlled by the  correlation length $\xi$ of vortical flows measured in simulations with uniform activity (SI). At small wavelengths, vortices of size $\xi>\lambda$ transport material across adjacent high and low activity regions, mixing the sample and suppressing the buildup of a density contrast. Conversely, when $\lambda>4\xi$ the density contrast plateaus.  

To visualize the spatial organization of the flow, we plot the $y$-averaged speed, $\langle |v| \rangle_y$, across the sample (Fig.~\ref{fig:vortex}(a)). For $\lambda=250~\mu\mathrm{m}$, the speed is nearly uniform, indicating that the flow does not resolve the imposed sinusoidal pattern. For larger wavelength $\lambda=1260~\mu\mathrm{m}$, the spatially modulated speed is faster in the high-activity regions; the converse is true in the dark regions. The vorticity maps provide a complementary view of this wavelength-dependent flow organization (Fig.~\ref{fig:vortex}(b)). At $\lambda=250~\mu\mathrm{m}$, vortices span neighboring high- and low-activity regions, producing chaotic mixing that advects material across the spatial pattern. This is consistent with the nearly uniform speed profile. At larger wavelength $\lambda=1260~\mu\mathrm{m}$, strong vortical flows are localized in the high-activity regions, providing further evidence of the role of vortex size and velocity correlations in controlling the flow and density pattern.  

Finally, we note the suppression of density contrast with increasing chamber thickness is not captured by $2D$ effective hydrodynamics (Fig.~\ref{fig:thickness}). The reduction to the effective $2D$ model assumes a quasi-$2D$ geometry in which the microtubule concentration, in-plane velocity, and nematic order are approximately uniform in the direction normal to the confining plates, while microtubule motion is confined to the plane. The solvent dynamics can then be eliminated by averaging across the gap. These approximations become less controlled for thicker chambers, where a fully three-dimensional two-fluid description is ultimately required.

\subsection{A minimal advection-diffusion equation}
\label{sec:diffusion}
When the spatial modulation of the activity pattern is large compared to the length scale of active flows, the inhomogeneous concentration profile is captured by a minimal continuum model that assumes overdamped dynamics ($\eta=0$) and neglects alignment ($\mathbf{Q}=0$). This model also neglects correlations and local microtubule alignment driven by deviatoric active stress. In the absence of viscous dissipation, the microtubule current  $\phi\mathbf{v}=-\frac{1}{\zeta}\boldsymbol{\nabla}(p+p_a)$ can be immediately eliminated from Eq.\eqref{eq:mass} to obtain an advection-diffusion equation for the microtubule concentration 
\begin{equation}
\partial_t\phi=\boldsymbol{\nabla}\cdot\left[-\mathbf{v}_d(\mathbf{r})\phi+\mathcal{D}(\mathbf{r})\boldsymbol{\nabla}\phi\right]\;,
\label{eq:diffusion}
\end{equation}
where $\mathbf{v}_d(\mathbf{r})=\boldsymbol{\nabla}\alpha_B(\mathbf{r})/\zeta$ and $\mathcal{D}(\mathbf{r})=D_p-\alpha_B(\mathbf{r})/\zeta$, with $D_p=D+(\kappa\zeta)^{-1}$. The effective advection diffusion equation is analogous to models that describe chemotaxis of bacteria with spatially dependent run speed and/or the tumble rate (Eq.~\eqref{eq:diffusion}) ~\cite{schnitzer1993theory,solon2018generalized,fily2012athermal,marchetti2016minimal}.  

The effective equation can be solved analytically for arbitrary one-dimensional illumination profile, $\alpha_B(x)=-(\alpha_0/3)f(x)$, with the result
\begin{equation}
\phi(x)=\frac{C}{1+R f(x)}\;,
\label{eq:1d-rho}
\end{equation}
where $R=D_a/D_p$, $D_a=\alpha_0/(3\zeta)$ is an active diffusivity and the constant $C$ is determined by mass conservation and the activity profile $f(x)$. For a step-wise activity profile $f(x)=\Theta(w/2-x)\Theta(x+w/2)$, with $\Theta(x)$ the Heaviside step function, the contrast density is given by $\rho_a/\rho_p=\frac{D_p}{D_p+D_a}$ which vanishes at high activity, and where $\rho_{a}$ and $\rho_{p}$ are respectively the densities of illuminated and dark regions (SI).

To compare with experiments, we examine again the sinusoidal activity profile, which gives $C=\sqrt{1+R}$. The resulting steady-state density contrast is wavelength independent; it magnitude is $A_1 = R/2\sqrt{1+R}$. The minimal model quantitatively captures the experimental density contrast (Fig.~\ref{fig:a_1}(a)) when the activity wavelength exceed the correlation length of active flows. It fails, however, to account for the suppression of the density modulation observed at small wavelengths (Fig.~\ref{fig:a_1}(a)). The values of $R$ close to unity used in the fits suggest that the experiments are in a regime where active and passive contributions to the filament diffusivities are comparable. The effective 1D model also explains the density profile asymmetry and its dependence on the activity profile sharpness (Fig.~\ref{fig:asymm}(c)). When solved for the hyperbolic tangent profile used in experiments, the model captures the dependence of asymmetry on the sharpness of imposed gradients (dashed line in Fig.~\ref{fig:asymm}(c)).

\section{Discussion}
We show that spatially patterned activity drives accumulation of microtubules into low activity regions, generating a density contrast that grows with the imposed wavelength and then saturates. At short wavelengths, vortices connect adjacent high- and low-activity regions to homogenize the suspension, whereas at larger wavelengths the flow becomes confined to regions of regions of different activity. In this limit, the behavior is qualitatively described by a minimal advection-diffusion equation that neglects hydrodynamic correlations driven by deviatoric active stresses. The sharpness of the illumination pattern controls the density modulation profile.

Many active fluids, from microtubule--kinesin mixtures to bacterial suspensions, are two-component systems: the active agents form a compressible medium within a solvent. The density, flow and alignment of the active agents are all spatially and temporally modulated. Yet previous theoretical work has mostly treated these systems as one-component fluids in one of two limits. On one hand, dry active matter models allow for density variation and compressibility, but neglect momentum-conserving flows~\cite{toner1995long,cates2015motility,maryshev2019dry,chate2022dry}. On the other hand, with few exceptions~\cite{doostmohammadi2016stabilization,bhattacharyya2023phase,bhattacharyya2025active}, wet active nematic theories incorporate hydrodynamic flows but typically assume incompressibility~\cite{giomi2014defect,giomi2015geometry, marenduzzo2007steady,shankar2024design}. 

We propose an effective continuum one-fluid model that incorporates flow, density variations and alignment, allowing them to dynamically influence each other. We show that isotropic active stresses compete with passive compressibility to push microtubule bundles out of the high-activity regions and accumulate them in the low activity regions. Isotropic active stresses were explored previously~\cite{marchetti2013hydrodynamics}, but mainly in the context of contractile active gels~\cite{prost2015active}. Their potential for designing density patterns in extensile active fluids has not been previously mapped out. Well-studied deviatoric active stresses drive chaotic flows. Activity modulation allows the simultaneous tuning of density and flow, providing a mechanism for control of the ubiquitous spatio-temporally chaotic dynamics of active fluids. 

The model presented here applies to quasi-2D fluids confined to thin chambers, where one can eliminate the solvent velocity by assuming that microtubule dynamics is effectively confined to the large plane of the chamber~\cite{krommydas2026controlling}. Confinement is known to strongly reorganize collective motion and active-nematic dynamics~\cite{Wioland2013,Wu2017,Nishiguchi2018,Hardouin2019,Opathalage2019,Chandrakar2020}. Experiments suggest that density accumulation is suppressed in thick chambers (Fig.~\ref{fig:thickness}), an effect that cannot be captured by our $2D$ model. A full two-fluid description will be needed to capture the behavior of bulk three dimensional systems. 

The accumulation of motile agents in low activity regions has been observed with light-sensitive \emph{E. coli}~\cite{arlt2018painting,frangipane2018dynamic}. In these systems, as in the one studied here, activity was modulated via external cues. In contrast to the microtubule-based system , however, the flow of bacterial suspensions was overdamped. Thus, the density accumulation in low motility region could be described using single-particle dynamics and the length scale of the density contrast was limited to the size and responsiveness of the microswimmers~\cite{Buttinoni2013,maggi2018currents}. At the continuum level the dynamics of light-responsive bacteria is captured by an advection-diffusion equation with spatially dependent motility parameters~\cite{schnitzer1993theory,cates2015motility,zhao2023chemotactic,bai2021spatial,TailleurCates2008, Stenhammar2016}. We showed that a similar scalar model with spatially-dependent activity captures density patterns in microtubule-based active fluids on scales large compared to the correlation length of active flows. A key distinction of our work, however, is that activity simultaneously modulates density and flow, in contract to previously studied dry systems. Both light responsive \emph{E. coli} and microtubules-based active fluids are distinct from motility-induced phase separation where motility is suppressed by intrinsic repulsive interactions among acting agents, and the resulting phase separation or density modulation is emergent.


\appendix
\section*{Materials and Methods}

\textbf{Continuum model and simulations.}
The continuum simulations use the quasi-two-dimensional compressible active-nematic model defined in the main text and detailed in the Supporting Information. The numerical results in the main text were obtained from two-dimensional simulations using our \texttt{cuPSS}-based pseudospectral implementation~\cite{krommydas2026compressible}. Simulations were performed on periodic square domains. The production runs used for the continuum comparison were performed with \(N_x=N_y=1024\), \(\Delta x=\Delta y=0.5\), and \(\Delta t=1\times10^{-4}\). Unless otherwise stated, the fixed continuum parameters were \(D=5.0\), \(K=1.0\), \(r=1.0\), \(\phi_{\rm IN}=1.43\), \(\langle\phi\rangle=1.0\), \(\kappa=0.476\), \(\eta=1.0\), \(\lambda_N=1.0\), \(\gamma=1.0\), \(u=5.0\), and \(\zeta=7\times10^{-5}\). These passive material parameters were kept fixed for all continuum simulations shown in the main wavelength-dependent comparison.

The large-amplitude continuum comparison used \((\alpha,\alpha_B)=(-7.7,-2.45)\), corresponding to \(R=1.17\) in the one-dimensional active-pressure model. The reduced-amplitude comparison used \((\alpha,\alpha_B)=(-4.62,-1.47)\), corresponding to \(R=0.70\). Thus both active coefficients were reduced by the same factor, \(0.6\), while all passive material parameters were kept fixed. This mirrors the experimental comparison between the two illumination amplitudes.

For the sinusoidal wavelength-dependent simulations, the imposed activity profile was \(f(x)=[1+\sin(2\pi x/\lambda)]/2\). The uniform-illumination simulations used to measure the intrinsic active-flow length scale had \(f=1\). The first-harmonic density contrast \(A_1\) was extracted from the \(y\)-averaged density profile at the imposed wavelength \(\lambda\). The intrinsic active-flow length scale \(\xi\) used to collapse the wavelength-dependent response was measured from uniform-activity simulations using the vorticity field. Specifically, we computed the shell-averaged vorticity structure factor \(S_\omega(q)\) and defined \(\xi=\ell_\omega\) as its inverse first moment.

To interpret the large-wavelength density contrast, we also used a minimal one-dimensional active-pressure diffusion model. This model is not an asymptotic solution of the full hydrodynamic equations. Rather, it is an approximation that isolates the isotropic active-pressure mechanism by neglecting nematic order, deviatoric active stresses, and vortical flow. In this approximation, the steady-state density profile for an imposed one-dimensional activity profile \(f(x)\) is \(\phi(x)=C/[1+Rf(x)]\), where \(C\) enforces conservation of microtubule number and \(R\) compares active-pressure-driven transport to passive diffusive and osmotic relaxation. This model captures the wavelength-independent large-\(\lambda\) plateau of the density contrast, but not the suppression of density contrast at small wavelengths, which requires the finite active-flow length scale of the full hydrodynamic model.

\textbf{Optically responsive microtubule-based active fluids.} Optically responsive microtubule-based active fluids were assembled using GMPCPP-stabilized microtubules, light-responsive kinesin motor constructs, an ATP-regeneration system, a depletion agent, and an oxygen-scavenging system. All samples were prepared in a buffer containing $80~\mathrm{mM}$ PIPES, $1~\mathrm{mM}$ EGTA, and $2~\mathrm{mM}$ MgCl$_2$ (pH 6.8; M2B buffer). The MgCl$_2$concentration was then increased to a final concentration of 5 mM. Tubulin was purified from bovine brain and fluorescently labeled following established protocols \cite{tayar2022assembling}. Approximately 3$\%$ of the tubulin was labeled with Alexa Fluor 647 NHS Ester (Invitrogen). Labeled and unlabeled tubulin were then co-polymerized into GMPCPP-stabilized microtubules in a M2B buffer. The motor system consisted of two complementary single-headed kinesin fragments fused to optically responsive iLID and micro domains. In the absence of blue light, the motor fragments remained predominantly unclustered and generated negligible active flows. Upon blue-light illumination, the iLID--micro interaction induced motor clustering, allowing kinesin clusters to bind neighboring microtubules and drive extensile interfilament sliding. The final active-fluid mixture contained microtubules at $0.3~\mathrm{mg/mL}$, micro-kinesin at $750~\mathrm{nM}$, iLID-kinesin at $750~\mathrm{nM}$, ATP at $2.2~\mathrm{mM}$, phosphoenol pyruvate (PEP) at $30~\mathrm{mM}$, pyruvate kinase/lactic dehydrogenase (PK/LDH) at approximately $30~\mathrm{units/mL}$, and $100~\mathrm{kDa}$ dextran as the depletion agent at $1\%$ w/v. The oxygen-scavenging system consisted of glucose at $3.7~\mathrm{mg/mL}$, glucose oxidase at $0.25~\mathrm{mg/mL}$, catalase at $0.043~\mathrm{mg/mL}$, and DTT at $6~\mu\mathrm{M}$. Microtubules were flash-frozen in aliquots and thawed immediately before sample preparation.

\textbf{Sample chambers.} Samples were confined in quasi--two-dimensional chambers assembled from acrylamide-brush-coated glass slides and coverslips~\cite{lau2009condensation}. The acrylamide brush passivated the glass surfaces and suppressed nonspecific adsorption of microtubules and protein components, which was especially important because the iLID-kinesin constructs were prone to adsorption on poorly coated glass surfaces. The chamber height was set by opaque white medical-grade adhesive tape spacers (Solventum/3M; product numbers 9966 and 9972A). The 9966 tape had a measured thickness of approximately $95~\mu\mathrm{m}$ and was used for $100~\mu\mathrm{m}$ chambers, while the 9972A tape had a measured thickness of approximately $190~\mu\mathrm{m}$ and was used for $200~\mu\mathrm{m}$ chambers. Taller chambers were made by stacking multiple spacer layers. Opaque spacers were used to minimize leakage of blue activation light from the chamber edges. Transparent spacers, such as parafilm, were avoided because they could guide or scatter blue light along the chamber boundary and distort the imposed illumination pattern. After loading the active-fluid mixture, chambers were sealed to prevent evaporation and mounted on the microscope for imaging.

\textbf{Optical activation and spatial light patterning.} Spatially patterned activity was imposed using blue activation light projected from above onto the sample through a digital micromirror device (DMD; DLP LightCrafter 4500, Texas Instruments). The DMD contains a $912 \times 1140$ array of micromirrors. Neighboring micromirrors are separated by $7.6~\mu\mathrm{m}$ on the DMD chip, and each micromirror was projected to approximately $10~\mu\mathrm{m}$ on the sample plane with a projection lens xx mm, as determined from calibration images. The full projected field was therefore approximately $11.5~\mathrm{mm} \times 9.2~\mathrm{mm}$. To minimize illumination nonuniformity and optical aberrations near the edge of the projected field, experiments were restricted to the central region of the projection, corresponding to approximately $50\%$ of the total projected area, or about $8.0~\mathrm{mm} \times 6.0~\mathrm{mm}$ on the sample plane. The projected illumination was further corrected using a flat-field calibration. The uniform pattern was projected onto a coverslip in the absence of a sample. Then the light pattern collected by an objective lens were recorded by the same camera equipped with a microscope. The desired activation patterns were then multiplied by the inverse of this measured intensity profile, producing a corrected input pattern that compensated for spatial variations in illumination intensity. This inverse-intensity correction flattened the projected blue-light field over the central region used for experiments. Before applying spatially patterned illumination, samples were uniformly activated for approximately $10~\mathrm{min}$ to homogenize the microtubule density and establish a reproducible initial state. The depth of focus of the projected illumination was approximately $1~\mathrm{mm}$, which was much larger than the chamber height; therefore, the imposed activation profile was nearly uniform across the sample thickness.

\textbf{Fluorescence imaging.} Microtubules were imaged by epi-fluorescence microscopy in the red channel using a $4\times$, 0.13 NA Nikon objective, while the blue channel was used for optical activation. The red excitation wavelength was spectrally separated from the blue activation light, and control experiments confirmed that red-channel imaging alone did not activate the opto-kinesin system or generate measurable active flows. The effective depth of field of the $4\times$ objective was approximately $60~\mu\mathrm{m}$, and the focal plane was placed near the middle of the chamber. Because the sample was confined in a thin quasi--two-dimensional chamber and imaged in wide-field fluorescence, the recorded red-channel intensity provided a projected fluorescence signal from microtubules over the chamber thickness. We therefore interpreted the images as a thickness-averaged, quasi--two-dimensional microtubule density field. Time-lapse image sequences were acquired after the onset of the projected light pattern to monitor the development of microtubule density modulation and active flows. For each experiment, the initial time $t=0$ was defined as the time at which the spatial activation pattern was applied.

\textbf{Density-profile analysis.} Microtubule density was estimated from the fluorescence intensity of the red-channel images. Images were acquired with an exposure time of $100~\mathrm{ms}$, and the camera histogram was checked to ensure that the fluorescence signal did not saturate the camera. Background images were acquired with the same exposure time to measure the camera and environmental background signal. The background intensity was approximately $700$ analog-to-digital units (ADU), for our imaging conditions.

For each raw fluorescence image $I_{\mathrm{raw}}(x,y,t)$, we first subtracted the background intensity $I_{\mathrm{bg}}$ from every pixel and then normalized by the mean background-subtracted intensity of the full image. The normalized microtubule density field was therefore defined as
\begin{equation}
\rho(x,y,t)=
\frac{
I_{\mathrm{raw}}(x,y,t)-I_{\mathrm{bg}}
}{
\left\langle I_{\mathrm{raw}}(x,y,t)\right\rangle_{x,y}-I_{\mathrm{bg}}
},
\end{equation}
where $\langle \cdot \rangle_{x,y}$ denotes an average over the full field of view. This normalization removes variations in total fluorescence intensity between experiments and gives a dimensionless density in arbitrary units, with spatial mean close to unity.

\textbf{Velocity-field analysis.} Microtubule velocity fields were measured from time-lapse fluorescence images using particle image velocimetry (PIV) in MATLAB with PIVlab. Consecutive images separated by $\Delta t=1~\mathrm{s}$ were analyzed using three passes with interrogation areas of $64$, $32$, and $16$ pixels and corresponding step sizes of $32$, $16$, and $8$ pixels. The resulting velocity field $\mathbf{v}(x,y,t)$ was used to calculate local speeds and spatially averaged flow profiles.

The intrinsic flow correlation length $\xi$ was measured under spatially uniform illumination. For each frame, the spatially averaged velocity was subtracted to obtain $\delta\mathbf{v}(\mathbf{r},t)=\mathbf{v}(\mathbf{r},t)-\langle\mathbf{v}\rangle_{\mathbf r}$. We then calculated the equal-time spatial autocorrelation of $\delta\mathbf{v}$, azimuthally averaged it over displacement directions, and averaged the resulting correlation function over time. The correlation length $\xi$ was defined as the first separation at which the normalized velocity autocorrelation decreased to $e^{-1}$ (Fig.~S1).

\bibliography{Biblio.bib}

\begin{thebibliography}{60}%
\makeatletter
\providecommand \@ifxundefined [1]{%
 \@ifx{#1\undefined}
}%
\providecommand \@ifnum [1]{%
 \ifnum #1\expandafter \@firstoftwo
 \else \expandafter \@secondoftwo
 \fi
}%
\providecommand \@ifx [1]{%
 \ifx #1\expandafter \@firstoftwo
 \else \expandafter \@secondoftwo
 \fi
}%
\providecommand \natexlab [1]{#1}%
\providecommand \enquote  [1]{``#1''}%
\providecommand \bibnamefont  [1]{#1}%
\providecommand \bibfnamefont [1]{#1}%
\providecommand \citenamefont [1]{#1}%
\providecommand \href@noop [0]{\@secondoftwo}%
\providecommand \href [0]{\begingroup \@sanitize@url \@href}%
\providecommand \@href[1]{\@@startlink{#1}\@@href}%
\providecommand \@@href[1]{\endgroup#1\@@endlink}%
\providecommand \@sanitize@url [0]{\catcode `\\12\catcode `\$12\catcode
  `\&12\catcode `\#12\catcode `\^12\catcode `\_12\catcode `\%12\relax}%
\providecommand \@@startlink[1]{}%
\providecommand \@@endlink[0]{}%
\providecommand \url  [0]{\begingroup\@sanitize@url \@url }%
\providecommand \@url [1]{\endgroup\@href {#1}{\urlprefix }}%
\providecommand \urlprefix  [0]{URL }%
\providecommand \Eprint [0]{\href }%
\providecommand \doibase [0]{https://doi.org/}%
\providecommand \selectlanguage [0]{\@gobble}%
\providecommand \bibinfo  [0]{\@secondoftwo}%
\providecommand \bibfield  [0]{\@secondoftwo}%
\providecommand \translation [1]{[#1]}%
\providecommand \BibitemOpen [0]{}%
\providecommand \bibitemStop [0]{}%
\providecommand \bibitemNoStop [0]{.\EOS\space}%
\providecommand \EOS [0]{\spacefactor3000\relax}%
\providecommand \BibitemShut  [1]{\csname bibitem#1\endcsname}%
\let\auto@bib@innerbib\@empty
\bibitem [{\citenamefont {Aditi~Simha}\ and\ \citenamefont
  {Ramaswamy}(2002)}]{aditi2002hydrodynamic}%
  \BibitemOpen
  \bibfield  {author} {\bibinfo {author} {\bibfnamefont {R.}~\bibnamefont
  {Aditi~Simha}}\ and\ \bibinfo {author} {\bibfnamefont {S.}~\bibnamefont
  {Ramaswamy}},\ }\bibfield  {title} {\bibinfo {title} {Hydrodynamic
  fluctuations and instabilities in ordered suspensions of self-propelled
  particles},\ }\href {https://doi.org/10.1103/PhysRevLett.89.058101}
  {\bibfield  {journal} {\bibinfo  {journal} {Phys. Rev. Lett.}\ }\textbf
  {\bibinfo {volume} {89}},\ \bibinfo {pages} {058101} (\bibinfo {year}
  {2002})}\BibitemShut {NoStop}%
\bibitem [{\citenamefont {Saintillan}\ and\ \citenamefont
  {Shelley}(2007)}]{saintillan2007orientational}%
  \BibitemOpen
  \bibfield  {author} {\bibinfo {author} {\bibfnamefont {D.}~\bibnamefont
  {Saintillan}}\ and\ \bibinfo {author} {\bibfnamefont {M.~J.}\ \bibnamefont
  {Shelley}},\ }\bibfield  {title} {\bibinfo {title} {Orientational order and
  instabilities in suspensions of self-locomoting rods},\ }\href
  {https://doi.org/10.1103/PhysRevLett.99.058102} {\bibfield  {journal}
  {\bibinfo  {journal} {Phys. Rev. Lett.}\ }\textbf {\bibinfo {volume} {99}},\
  \bibinfo {pages} {058102} (\bibinfo {year} {2007})}\BibitemShut {NoStop}%
\bibitem [{\citenamefont {Sanchez}\ \emph {et~al.}(2012)\citenamefont
  {Sanchez}, \citenamefont {Chen}, \citenamefont {DeCamp}, \citenamefont
  {Heymann},\ and\ \citenamefont {Dogic}}]{sanchez2012spontaneous}%
  \BibitemOpen
  \bibfield  {author} {\bibinfo {author} {\bibfnamefont {T.}~\bibnamefont
  {Sanchez}}, \bibinfo {author} {\bibfnamefont {D.~T.~N.}\ \bibnamefont
  {Chen}}, \bibinfo {author} {\bibfnamefont {S.~J.}\ \bibnamefont {DeCamp}},
  \bibinfo {author} {\bibfnamefont {M.}~\bibnamefont {Heymann}},\ and\ \bibinfo
  {author} {\bibfnamefont {Z.}~\bibnamefont {Dogic}},\ }\bibfield  {title}
  {\bibinfo {title} {Spontaneous motion in hierarchically assembled active
  matter},\ }\href {https://doi.org/10.1038/nature11591} {\bibfield  {journal}
  {\bibinfo  {journal} {Nature}\ }\textbf {\bibinfo {volume} {491}},\ \bibinfo
  {pages} {431} (\bibinfo {year} {2012})}\BibitemShut {NoStop}%
\bibitem [{\citenamefont {Kumar}\ \emph {et~al.}(2018)\citenamefont {Kumar},
  \citenamefont {Zhang}, \citenamefont {De~Pablo},\ and\ \citenamefont
  {Gardel}}]{kumar2018tunable}%
  \BibitemOpen
  \bibfield  {author} {\bibinfo {author} {\bibfnamefont {N.}~\bibnamefont
  {Kumar}}, \bibinfo {author} {\bibfnamefont {R.}~\bibnamefont {Zhang}},
  \bibinfo {author} {\bibfnamefont {J.~J.}\ \bibnamefont {De~Pablo}},\ and\
  \bibinfo {author} {\bibfnamefont {M.~L.}\ \bibnamefont {Gardel}},\ }\bibfield
   {title} {\bibinfo {title} {Tunable structure and dynamics of active liquid
  crystals},\ }\href@noop {} {\bibfield  {journal} {\bibinfo  {journal}
  {Science advances}\ }\textbf {\bibinfo {volume} {4}},\ \bibinfo {pages}
  {eaat7779} (\bibinfo {year} {2018})}\BibitemShut {NoStop}%
\bibitem [{\citenamefont {Wioland}\ \emph
  {et~al.}(2013{\natexlab{a}})\citenamefont {Wioland}, \citenamefont
  {Woodhouse}, \citenamefont {Dunkel}, \citenamefont {Kessler},\ and\
  \citenamefont {Goldstein}}]{wioland2013confinement}%
  \BibitemOpen
  \bibfield  {author} {\bibinfo {author} {\bibfnamefont {H.}~\bibnamefont
  {Wioland}}, \bibinfo {author} {\bibfnamefont {F.~G.}\ \bibnamefont
  {Woodhouse}}, \bibinfo {author} {\bibfnamefont {J.}~\bibnamefont {Dunkel}},
  \bibinfo {author} {\bibfnamefont {J.~O.}\ \bibnamefont {Kessler}},\ and\
  \bibinfo {author} {\bibfnamefont {R.~E.}\ \bibnamefont {Goldstein}},\
  }\bibfield  {title} {\bibinfo {title} {Confinement stabilizes a bacterial
  suspension into a spiral vortex},\ }\href@noop {} {\bibfield  {journal}
  {\bibinfo  {journal} {Physical review letters}\ }\textbf {\bibinfo {volume}
  {110}},\ \bibinfo {pages} {268102} (\bibinfo {year}
  {2013}{\natexlab{a}})}\BibitemShut {NoStop}%
\bibitem [{\citenamefont {Wu}\ \emph {et~al.}(2017{\natexlab{a}})\citenamefont
  {Wu}, \citenamefont {Hishamunda}, \citenamefont {Chen}, \citenamefont
  {DeCamp}, \citenamefont {Chang}, \citenamefont {Fern{\'a}ndez-Nieves},
  \citenamefont {Fraden},\ and\ \citenamefont {Dogic}}]{wu2017transition}%
  \BibitemOpen
  \bibfield  {author} {\bibinfo {author} {\bibfnamefont {K.-T.}\ \bibnamefont
  {Wu}}, \bibinfo {author} {\bibfnamefont {J.~B.}\ \bibnamefont {Hishamunda}},
  \bibinfo {author} {\bibfnamefont {D.~T.}\ \bibnamefont {Chen}}, \bibinfo
  {author} {\bibfnamefont {S.~J.}\ \bibnamefont {DeCamp}}, \bibinfo {author}
  {\bibfnamefont {Y.-W.}\ \bibnamefont {Chang}}, \bibinfo {author}
  {\bibfnamefont {A.}~\bibnamefont {Fern{\'a}ndez-Nieves}}, \bibinfo {author}
  {\bibfnamefont {S.}~\bibnamefont {Fraden}},\ and\ \bibinfo {author}
  {\bibfnamefont {Z.}~\bibnamefont {Dogic}},\ }\bibfield  {title} {\bibinfo
  {title} {Transition from turbulent to coherent flows in confined
  three-dimensional active fluids},\ }\href@noop {} {\bibfield  {journal}
  {\bibinfo  {journal} {Science}\ }\textbf {\bibinfo {volume} {355}},\ \bibinfo
  {pages} {eaal1979} (\bibinfo {year} {2017}{\natexlab{a}})}\BibitemShut
  {NoStop}%
\bibitem [{\citenamefont {Brugu{\'e}s}\ and\ \citenamefont
  {Needleman}(2014)}]{brugues2014physical}%
  \BibitemOpen
  \bibfield  {author} {\bibinfo {author} {\bibfnamefont {J.}~\bibnamefont
  {Brugu{\'e}s}}\ and\ \bibinfo {author} {\bibfnamefont {D.}~\bibnamefont
  {Needleman}},\ }\bibfield  {title} {\bibinfo {title} {Physical basis of
  spindle self-organization},\ }\href {https://doi.org/10.1073/pnas.1409404111}
  {\bibfield  {journal} {\bibinfo  {journal} {Proc. Natl. Acad. Sci. U.S.A.}\
  }\textbf {\bibinfo {volume} {111}},\ \bibinfo {pages} {18496} (\bibinfo
  {year} {2014})}\BibitemShut {NoStop}%
\bibitem [{\citenamefont {Mayer}\ \emph {et~al.}(2010)\citenamefont {Mayer},
  \citenamefont {Depken}, \citenamefont {Bois}, \citenamefont {J{\"u}licher},\
  and\ \citenamefont {Grill}}]{mayer2010anisotropies}%
  \BibitemOpen
  \bibfield  {author} {\bibinfo {author} {\bibfnamefont {M.}~\bibnamefont
  {Mayer}}, \bibinfo {author} {\bibfnamefont {M.}~\bibnamefont {Depken}},
  \bibinfo {author} {\bibfnamefont {J.~S.}\ \bibnamefont {Bois}}, \bibinfo
  {author} {\bibfnamefont {F.}~\bibnamefont {J{\"u}licher}},\ and\ \bibinfo
  {author} {\bibfnamefont {S.~W.}\ \bibnamefont {Grill}},\ }\bibfield  {title}
  {\bibinfo {title} {Anisotropies in cortical tension reveal the physical basis
  of polarizing cortical flows},\ }\href {https://doi.org/10.1038/nature09376}
  {\bibfield  {journal} {\bibinfo  {journal} {Nature}\ }\textbf {\bibinfo
  {volume} {467}},\ \bibinfo {pages} {617} (\bibinfo {year}
  {2010})}\BibitemShut {NoStop}%
\bibitem [{\citenamefont {Mitchell}\ \emph {et~al.}(2022)\citenamefont
  {Mitchell}, \citenamefont {Cislo}, \citenamefont {Shankar}, \citenamefont
  {Lin}, \citenamefont {Shraiman},\ and\ \citenamefont
  {Streichan}}]{mitchell2022visceral}%
  \BibitemOpen
  \bibfield  {author} {\bibinfo {author} {\bibfnamefont {N.~P.}\ \bibnamefont
  {Mitchell}}, \bibinfo {author} {\bibfnamefont {D.~J.}\ \bibnamefont {Cislo}},
  \bibinfo {author} {\bibfnamefont {S.}~\bibnamefont {Shankar}}, \bibinfo
  {author} {\bibfnamefont {Y.}~\bibnamefont {Lin}}, \bibinfo {author}
  {\bibfnamefont {B.~I.}\ \bibnamefont {Shraiman}},\ and\ \bibinfo {author}
  {\bibfnamefont {S.~J.}\ \bibnamefont {Streichan}},\ }\bibfield  {title}
  {\bibinfo {title} {Visceral organ morphogenesis via calcium-patterned muscle
  constrictions},\ }\href {https://doi.org/10.7554/eLife.77355} {\bibfield
  {journal} {\bibinfo  {journal} {eLife}\ }\textbf {\bibinfo {volume} {11}},\
  \bibinfo {pages} {e77355} (\bibinfo {year} {2022})}\BibitemShut {NoStop}%
\bibitem [{\citenamefont {Guglielmi}\ \emph {et~al.}(2015)\citenamefont
  {Guglielmi}, \citenamefont {Barry}, \citenamefont {Huber},\ and\
  \citenamefont {De~Renzis}}]{Guglielmi2015}%
  \BibitemOpen
  \bibfield  {author} {\bibinfo {author} {\bibfnamefont {G.}~\bibnamefont
  {Guglielmi}}, \bibinfo {author} {\bibfnamefont {J.~D.}\ \bibnamefont
  {Barry}}, \bibinfo {author} {\bibfnamefont {W.}~\bibnamefont {Huber}},\ and\
  \bibinfo {author} {\bibfnamefont {S.}~\bibnamefont {De~Renzis}},\ }\bibfield
  {title} {\bibinfo {title} {An optogenetic method to modulate cell
  contractility during tissue morphogenesis},\ }\href
  {https://doi.org/10.1016/j.devcel.2015.10.020} {\bibfield  {journal}
  {\bibinfo  {journal} {Dev. Cell}\ }\textbf {\bibinfo {volume} {35}},\
  \bibinfo {pages} {646} (\bibinfo {year} {2015})}\BibitemShut {NoStop}%
\bibitem [{\citenamefont {Ross}\ \emph {et~al.}(2019)\citenamefont {Ross},
  \citenamefont {Lee}, \citenamefont {Qu}, \citenamefont {Banks}, \citenamefont
  {Phillips},\ and\ \citenamefont {Thomson}}]{Ross2019Nature}%
  \BibitemOpen
  \bibfield  {author} {\bibinfo {author} {\bibfnamefont {T.~D.}\ \bibnamefont
  {Ross}}, \bibinfo {author} {\bibfnamefont {H.~J.}\ \bibnamefont {Lee}},
  \bibinfo {author} {\bibfnamefont {Z.}~\bibnamefont {Qu}}, \bibinfo {author}
  {\bibfnamefont {R.~A.}\ \bibnamefont {Banks}}, \bibinfo {author}
  {\bibfnamefont {R.}~\bibnamefont {Phillips}},\ and\ \bibinfo {author}
  {\bibfnamefont {M.}~\bibnamefont {Thomson}},\ }\bibfield  {title} {\bibinfo
  {title} {Controlling organization and forces in active matter through
  optically defined boundaries},\ }\href
  {https://doi.org/10.1038/s41586-019-1447-1} {\bibfield  {journal} {\bibinfo
  {journal} {Nature}\ }\textbf {\bibinfo {volume} {572}},\ \bibinfo {pages}
  {224} (\bibinfo {year} {2019})}\BibitemShut {NoStop}%
\bibitem [{\citenamefont {Peng}\ \emph {et~al.}(2021)\citenamefont {Peng},
  \citenamefont {Liu},\ and\ \citenamefont {Cheng}}]{peng2021imaging}%
  \BibitemOpen
  \bibfield  {author} {\bibinfo {author} {\bibfnamefont {Y.}~\bibnamefont
  {Peng}}, \bibinfo {author} {\bibfnamefont {Z.}~\bibnamefont {Liu}},\ and\
  \bibinfo {author} {\bibfnamefont {X.}~\bibnamefont {Cheng}},\ }\bibfield
  {title} {\bibinfo {title} {Imaging the emergence of bacterial turbulence:
  Phase diagram and transition kinetics},\ }\href@noop {} {\bibfield  {journal}
  {\bibinfo  {journal} {Science advances}\ }\textbf {\bibinfo {volume} {7}},\
  \bibinfo {pages} {eabd1240} (\bibinfo {year} {2021})}\BibitemShut {NoStop}%
\bibitem [{\citenamefont {Yang}\ \emph {et~al.}(2025)\citenamefont {Yang},
  \citenamefont {Liu}, \citenamefont {Lee}, \citenamefont {Phillips},\ and\
  \citenamefont {Thomson}}]{yang2025dynamic}%
  \BibitemOpen
  \bibfield  {author} {\bibinfo {author} {\bibfnamefont {F.}~\bibnamefont
  {Yang}}, \bibinfo {author} {\bibfnamefont {S.}~\bibnamefont {Liu}}, \bibinfo
  {author} {\bibfnamefont {H.~J.}\ \bibnamefont {Lee}}, \bibinfo {author}
  {\bibfnamefont {R.}~\bibnamefont {Phillips}},\ and\ \bibinfo {author}
  {\bibfnamefont {M.}~\bibnamefont {Thomson}},\ }\bibfield  {title} {\bibinfo
  {title} {Dynamic flow control through active matter programming language},\
  }\href@noop {} {\bibfield  {journal} {\bibinfo  {journal} {Nature Materials}\
  }\textbf {\bibinfo {volume} {24}},\ \bibinfo {pages} {615} (\bibinfo {year}
  {2025})}\BibitemShut {NoStop}%
\bibitem [{\citenamefont {Lemma}\ \emph {et~al.}(2023)\citenamefont {Lemma},
  \citenamefont {Varghese}, \citenamefont {Ross}, \citenamefont {Thomson},
  \citenamefont {Baskaran},\ and\ \citenamefont {Dogic}}]{LinneaPNASnexus}%
  \BibitemOpen
  \bibfield  {author} {\bibinfo {author} {\bibfnamefont {L.~M.}\ \bibnamefont
  {Lemma}}, \bibinfo {author} {\bibfnamefont {M.}~\bibnamefont {Varghese}},
  \bibinfo {author} {\bibfnamefont {T.~D.}\ \bibnamefont {Ross}}, \bibinfo
  {author} {\bibfnamefont {M.}~\bibnamefont {Thomson}}, \bibinfo {author}
  {\bibfnamefont {A.}~\bibnamefont {Baskaran}},\ and\ \bibinfo {author}
  {\bibfnamefont {Z.}~\bibnamefont {Dogic}},\ }\bibfield  {title} {\bibinfo
  {title} {Spatio-temporal patterning of extensile active stresses in
  microtubule-based active fluids},\ }\href
  {https://doi.org/10.1093/pnasnexus/pgad130} {\bibfield  {journal} {\bibinfo
  {journal} {PNAS Nexus}\ }\textbf {\bibinfo {volume} {2}},\ \bibinfo {pages}
  {pgad130} (\bibinfo {year} {2023})}\BibitemShut {NoStop}%
\bibitem [{\citenamefont {Wensink}\ \emph {et~al.}(2012)\citenamefont
  {Wensink}, \citenamefont {Dunkel}, \citenamefont {Heidenreich}, \citenamefont
  {Drescher}, \citenamefont {Goldstein}, \citenamefont {L{\"o}wen},\ and\
  \citenamefont {Yeomans}}]{wensink2012meso}%
  \BibitemOpen
  \bibfield  {author} {\bibinfo {author} {\bibfnamefont {H.~H.}\ \bibnamefont
  {Wensink}}, \bibinfo {author} {\bibfnamefont {J.}~\bibnamefont {Dunkel}},
  \bibinfo {author} {\bibfnamefont {S.}~\bibnamefont {Heidenreich}}, \bibinfo
  {author} {\bibfnamefont {K.}~\bibnamefont {Drescher}}, \bibinfo {author}
  {\bibfnamefont {R.~E.}\ \bibnamefont {Goldstein}}, \bibinfo {author}
  {\bibfnamefont {H.}~\bibnamefont {L{\"o}wen}},\ and\ \bibinfo {author}
  {\bibfnamefont {J.~M.}\ \bibnamefont {Yeomans}},\ }\bibfield  {title}
  {\bibinfo {title} {Meso-scale turbulence in living fluids},\ }\href@noop {}
  {\bibfield  {journal} {\bibinfo  {journal} {Proceedings of the national
  academy of sciences}\ }\textbf {\bibinfo {volume} {109}},\ \bibinfo {pages}
  {14308} (\bibinfo {year} {2012})}\BibitemShut {NoStop}%
\bibitem [{\citenamefont {Zhou}\ \emph {et~al.}(2014)\citenamefont {Zhou},
  \citenamefont {Sokolov}, \citenamefont {Lavrentovich},\ and\ \citenamefont
  {Aranson}}]{zhou2014living}%
  \BibitemOpen
  \bibfield  {author} {\bibinfo {author} {\bibfnamefont {S.}~\bibnamefont
  {Zhou}}, \bibinfo {author} {\bibfnamefont {A.}~\bibnamefont {Sokolov}},
  \bibinfo {author} {\bibfnamefont {O.~D.}\ \bibnamefont {Lavrentovich}},\ and\
  \bibinfo {author} {\bibfnamefont {I.~S.}\ \bibnamefont {Aranson}},\
  }\bibfield  {title} {\bibinfo {title} {Living liquid crystals},\ }\href
  {https://doi.org/10.1016/j.bpj.2013.11.2363} {\bibfield  {journal} {\bibinfo
  {journal} {Biophys. J.}\ }\textbf {\bibinfo {volume} {106}},\ \bibinfo
  {pages} {420a} (\bibinfo {year} {2014})}\BibitemShut {NoStop}%
\bibitem [{\citenamefont {Marchetti}\ \emph {et~al.}(2013)\citenamefont
  {Marchetti}, \citenamefont {Joanny}, \citenamefont {Ramaswamy}, \citenamefont
  {Liverpool}, \citenamefont {Prost}, \citenamefont {Rao},\ and\ \citenamefont
  {Simha}}]{marchetti2013hydrodynamics}%
  \BibitemOpen
  \bibfield  {author} {\bibinfo {author} {\bibfnamefont {M.~C.}\ \bibnamefont
  {Marchetti}}, \bibinfo {author} {\bibfnamefont {J.-F.}\ \bibnamefont
  {Joanny}}, \bibinfo {author} {\bibfnamefont {S.}~\bibnamefont {Ramaswamy}},
  \bibinfo {author} {\bibfnamefont {T.~B.}\ \bibnamefont {Liverpool}}, \bibinfo
  {author} {\bibfnamefont {J.}~\bibnamefont {Prost}}, \bibinfo {author}
  {\bibfnamefont {M.}~\bibnamefont {Rao}},\ and\ \bibinfo {author}
  {\bibfnamefont {R.~A.}\ \bibnamefont {Simha}},\ }\bibfield  {title} {\bibinfo
  {title} {Hydrodynamics of soft active matter},\ }\href
  {https://doi.org/10.1103/RevModPhys.85.1143} {\bibfield  {journal} {\bibinfo
  {journal} {Rev. Mod. Phys.}\ }\textbf {\bibinfo {volume} {85}},\ \bibinfo
  {pages} {1143} (\bibinfo {year} {2013})}\BibitemShut {NoStop}%
\bibitem [{\citenamefont {Arlt}\ \emph {et~al.}(2018)\citenamefont {Arlt},
  \citenamefont {Martinez}, \citenamefont {Dawson}, \citenamefont {Pilizota},\
  and\ \citenamefont {Poon}}]{arlt2018painting}%
  \BibitemOpen
  \bibfield  {author} {\bibinfo {author} {\bibfnamefont {J.}~\bibnamefont
  {Arlt}}, \bibinfo {author} {\bibfnamefont {V.~A.}\ \bibnamefont {Martinez}},
  \bibinfo {author} {\bibfnamefont {A.}~\bibnamefont {Dawson}}, \bibinfo
  {author} {\bibfnamefont {T.}~\bibnamefont {Pilizota}},\ and\ \bibinfo
  {author} {\bibfnamefont {W.~C.~K.}\ \bibnamefont {Poon}},\ }\bibfield
  {title} {\bibinfo {title} {Painting with light-powered bacteria},\ }\href
  {https://doi.org/10.1038/s41467-018-03161-8} {\bibfield  {journal} {\bibinfo
  {journal} {Nat. Commun.}\ }\textbf {\bibinfo {volume} {9}},\ \bibinfo {pages}
  {768} (\bibinfo {year} {2018})}\BibitemShut {NoStop}%
\bibitem [{\citenamefont {Frangipane}\ \emph {et~al.}(2018)\citenamefont
  {Frangipane}, \citenamefont {Dell'Arciprete}, \citenamefont {Petracchini},
  \citenamefont {Maggi}, \citenamefont {Saglimbeni}, \citenamefont {Bianchi},
  \citenamefont {Vizsnyiczai}, \citenamefont {Bernardini},\ and\ \citenamefont
  {Di~Leonardo}}]{frangipane2018dynamic}%
  \BibitemOpen
  \bibfield  {author} {\bibinfo {author} {\bibfnamefont {G.}~\bibnamefont
  {Frangipane}}, \bibinfo {author} {\bibfnamefont {D.}~\bibnamefont
  {Dell'Arciprete}}, \bibinfo {author} {\bibfnamefont {S.}~\bibnamefont
  {Petracchini}}, \bibinfo {author} {\bibfnamefont {C.}~\bibnamefont {Maggi}},
  \bibinfo {author} {\bibfnamefont {F.}~\bibnamefont {Saglimbeni}}, \bibinfo
  {author} {\bibfnamefont {S.}~\bibnamefont {Bianchi}}, \bibinfo {author}
  {\bibfnamefont {G.}~\bibnamefont {Vizsnyiczai}}, \bibinfo {author}
  {\bibfnamefont {M.~L.}\ \bibnamefont {Bernardini}},\ and\ \bibinfo {author}
  {\bibfnamefont {R.}~\bibnamefont {Di~Leonardo}},\ }\bibfield  {title}
  {\bibinfo {title} {Dynamic density shaping of photokinetic {E. coli}},\
  }\href {https://doi.org/10.7554/eLife.36608} {\bibfield  {journal} {\bibinfo
  {journal} {eLife}\ }\textbf {\bibinfo {volume} {7}},\ \bibinfo {pages}
  {e36608} (\bibinfo {year} {2018})}\BibitemShut {NoStop}%
\bibitem [{\citenamefont {Krommydas}(2026)}]{krommydas2026compressible}%
  \BibitemOpen
  \bibfield  {author} {\bibinfo {author} {\bibfnamefont {D.}~\bibnamefont
  {Krommydas}},\ }\href@noop {} {\bibinfo {title} {Compressible active nematics
  with {cuPSS}}},\ \bibinfo {howpublished}
  {\url{https://github.com/KrommydasDi/compressible-active-nematics-cuPSS}}
  (\bibinfo {year} {2026}),\ \bibinfo {note} {{CUDA}-based simulation code for
  compressible active nematics}\BibitemShut {NoStop}%
\bibitem [{\citenamefont {Schuppler}\ \emph {et~al.}(2016)\citenamefont
  {Schuppler}, \citenamefont {Keber}, \citenamefont {Kr{\"o}ger},\ and\
  \citenamefont {Bausch}}]{schuppler2016boundaries}%
  \BibitemOpen
  \bibfield  {author} {\bibinfo {author} {\bibfnamefont {M.}~\bibnamefont
  {Schuppler}}, \bibinfo {author} {\bibfnamefont {F.~C.}\ \bibnamefont
  {Keber}}, \bibinfo {author} {\bibfnamefont {M.}~\bibnamefont {Kr{\"o}ger}},\
  and\ \bibinfo {author} {\bibfnamefont {A.~R.}\ \bibnamefont {Bausch}},\
  }\bibfield  {title} {\bibinfo {title} {Boundaries steer the contraction of
  active gels},\ }\href {https://doi.org/10.1038/ncomms13120} {\bibfield
  {journal} {\bibinfo  {journal} {Nat. Commun.}\ }\textbf {\bibinfo {volume}
  {7}},\ \bibinfo {pages} {13120} (\bibinfo {year} {2016})}\BibitemShut
  {NoStop}%
\bibitem [{\citenamefont {Linsmeier}\ \emph {et~al.}(2016)\citenamefont
  {Linsmeier}, \citenamefont {Banerjee}, \citenamefont {Oakes}, \citenamefont
  {Jung}, \citenamefont {Kim},\ and\ \citenamefont
  {Murrell}}]{linsmeier2016disordered}%
  \BibitemOpen
  \bibfield  {author} {\bibinfo {author} {\bibfnamefont {I.}~\bibnamefont
  {Linsmeier}}, \bibinfo {author} {\bibfnamefont {S.}~\bibnamefont {Banerjee}},
  \bibinfo {author} {\bibfnamefont {P.~W.}\ \bibnamefont {Oakes}}, \bibinfo
  {author} {\bibfnamefont {W.}~\bibnamefont {Jung}}, \bibinfo {author}
  {\bibfnamefont {T.}~\bibnamefont {Kim}},\ and\ \bibinfo {author}
  {\bibfnamefont {M.~P.}\ \bibnamefont {Murrell}},\ }\bibfield  {title}
  {\bibinfo {title} {Disordered actomyosin networks are sufficient to produce
  cooperative and telescopic contractility},\ }\href
  {https://doi.org/10.1038/ncomms12615} {\bibfield  {journal} {\bibinfo
  {journal} {Nat. Commun.}\ }\textbf {\bibinfo {volume} {7}},\ \bibinfo {pages}
  {12615} (\bibinfo {year} {2016})}\BibitemShut {NoStop}%
\bibitem [{\citenamefont {Norton}\ \emph {et~al.}(2020)\citenamefont {Norton},
  \citenamefont {Grover}, \citenamefont {Hagan},\ and\ \citenamefont
  {Fraden}}]{norton2020optimal}%
  \BibitemOpen
  \bibfield  {author} {\bibinfo {author} {\bibfnamefont {M.~M.}\ \bibnamefont
  {Norton}}, \bibinfo {author} {\bibfnamefont {P.}~\bibnamefont {Grover}},
  \bibinfo {author} {\bibfnamefont {M.~F.}\ \bibnamefont {Hagan}},\ and\
  \bibinfo {author} {\bibfnamefont {S.}~\bibnamefont {Fraden}},\ }\bibfield
  {title} {\bibinfo {title} {Optimal control of active nematics},\ }\href
  {https://doi.org/10.1103/PhysRevLett.125.178005} {\bibfield  {journal}
  {\bibinfo  {journal} {Physical review letters}\ }\textbf {\bibinfo {volume}
  {125}},\ \bibinfo {pages} {178005} (\bibinfo {year} {2020})}\BibitemShut
  {NoStop}%
\bibitem [{\citenamefont {Zarei}\ \emph {et~al.}(2023)\citenamefont {Zarei},
  \citenamefont {Berezney}, \citenamefont {Hensley}, \citenamefont {Lemma},
  \citenamefont {Senbil}, \citenamefont {Dogic},\ and\ \citenamefont
  {Fraden}}]{zarei2023light}%
  \BibitemOpen
  \bibfield  {author} {\bibinfo {author} {\bibfnamefont {Z.}~\bibnamefont
  {Zarei}}, \bibinfo {author} {\bibfnamefont {J.}~\bibnamefont {Berezney}},
  \bibinfo {author} {\bibfnamefont {A.}~\bibnamefont {Hensley}}, \bibinfo
  {author} {\bibfnamefont {L.}~\bibnamefont {Lemma}}, \bibinfo {author}
  {\bibfnamefont {N.}~\bibnamefont {Senbil}}, \bibinfo {author} {\bibfnamefont
  {Z.}~\bibnamefont {Dogic}},\ and\ \bibinfo {author} {\bibfnamefont
  {S.}~\bibnamefont {Fraden}},\ }\bibfield  {title} {\bibinfo {title}
  {Light-activated microtubule-based two-dimensional active nematic},\ }\href
  {https://doi.org/10.1039/D3SM00270E} {\bibfield  {journal} {\bibinfo
  {journal} {Soft Matter}\ }\textbf {\bibinfo {volume} {19}},\ \bibinfo {pages}
  {6691} (\bibinfo {year} {2023})}\BibitemShut {NoStop}%
\bibitem [{\citenamefont {Shankar}\ \emph {et~al.}(2024)\citenamefont
  {Shankar}, \citenamefont {Scharrer}, \citenamefont {Bowick},\ and\
  \citenamefont {Marchetti}}]{shankar2024design}%
  \BibitemOpen
  \bibfield  {author} {\bibinfo {author} {\bibfnamefont {S.}~\bibnamefont
  {Shankar}}, \bibinfo {author} {\bibfnamefont {L.~V.~D.}\ \bibnamefont
  {Scharrer}}, \bibinfo {author} {\bibfnamefont {M.~J.}\ \bibnamefont
  {Bowick}},\ and\ \bibinfo {author} {\bibfnamefont {M.~C.}\ \bibnamefont
  {Marchetti}},\ }\bibfield  {title} {\bibinfo {title} {Design rules for
  controlling active topological defects},\ }\href
  {https://doi.org/10.1073/pnas.2400933121} {\bibfield  {journal} {\bibinfo
  {journal} {Proc. Natl. Acad. Sci. U.S.A.}\ }\textbf {\bibinfo {volume}
  {121}},\ \bibinfo {pages} {e2400933121} (\bibinfo {year} {2024})}\BibitemShut
  {NoStop}%
\bibitem [{\citenamefont {Nishiyama}\ \emph {et~al.}(2025)\citenamefont
  {Nishiyama}, \citenamefont {Berezney}, \citenamefont {Norton}, \citenamefont
  {Aggarwal}, \citenamefont {Ghosh}, \citenamefont {Zarei}, \citenamefont
  {Hagan}, \citenamefont {Fraden},\ and\ \citenamefont
  {Dogic}}]{nishiyama2025closed}%
  \BibitemOpen
  \bibfield  {author} {\bibinfo {author} {\bibfnamefont {K.}~\bibnamefont
  {Nishiyama}}, \bibinfo {author} {\bibfnamefont {J.}~\bibnamefont {Berezney}},
  \bibinfo {author} {\bibfnamefont {M.~M.}\ \bibnamefont {Norton}}, \bibinfo
  {author} {\bibfnamefont {A.}~\bibnamefont {Aggarwal}}, \bibinfo {author}
  {\bibfnamefont {S.}~\bibnamefont {Ghosh}}, \bibinfo {author} {\bibfnamefont
  {Z.}~\bibnamefont {Zarei}}, \bibinfo {author} {\bibfnamefont {M.~F.}\
  \bibnamefont {Hagan}}, \bibinfo {author} {\bibfnamefont {S.}~\bibnamefont
  {Fraden}},\ and\ \bibinfo {author} {\bibfnamefont {Z.}~\bibnamefont
  {Dogic}},\ }\bibfield  {title} {\bibinfo {title} {Closed-loop control of
  active nematic flows},\ }\href {https://doi.org/10.1103/4hrx-6rdq} {\bibfield
   {journal} {\bibinfo  {journal} {Physical Review X}\ }\textbf {\bibinfo
  {volume} {15}},\ \bibinfo {pages} {041053} (\bibinfo {year}
  {2025})}\BibitemShut {NoStop}%
\bibitem [{\citenamefont {Nedelec}\ \emph {et~al.}(1997)\citenamefont
  {Nedelec}, \citenamefont {Surrey}, \citenamefont {Maggs},\ and\ \citenamefont
  {Leibler}}]{nedelec1997self}%
  \BibitemOpen
  \bibfield  {author} {\bibinfo {author} {\bibfnamefont {F.}~\bibnamefont
  {Nedelec}}, \bibinfo {author} {\bibfnamefont {T.}~\bibnamefont {Surrey}},
  \bibinfo {author} {\bibfnamefont {A.~C.}\ \bibnamefont {Maggs}},\ and\
  \bibinfo {author} {\bibfnamefont {S.}~\bibnamefont {Leibler}},\ }\bibfield
  {title} {\bibinfo {title} {Self-organization of microtubules and motors},\
  }\href {https://doi.org/10.1038/38532} {\bibfield  {journal} {\bibinfo
  {journal} {Nature}\ }\textbf {\bibinfo {volume} {389}},\ \bibinfo {pages}
  {305} (\bibinfo {year} {1997})}\BibitemShut {NoStop}%
\bibitem [{\citenamefont {Santhosh}\ \emph {et~al.}(2020)\citenamefont
  {Santhosh}, \citenamefont {Nejad}, \citenamefont {Doostmohammadi},
  \citenamefont {Yeomans},\ and\ \citenamefont
  {Thampi}}]{santhosh2020activity}%
  \BibitemOpen
  \bibfield  {author} {\bibinfo {author} {\bibfnamefont {S.}~\bibnamefont
  {Santhosh}}, \bibinfo {author} {\bibfnamefont {M.~R.}\ \bibnamefont {Nejad}},
  \bibinfo {author} {\bibfnamefont {A.}~\bibnamefont {Doostmohammadi}},
  \bibinfo {author} {\bibfnamefont {J.~M.}\ \bibnamefont {Yeomans}},\ and\
  \bibinfo {author} {\bibfnamefont {S.~P.}\ \bibnamefont {Thampi}},\ }\bibfield
   {title} {\bibinfo {title} {Activity induced nematic order in isotropic
  liquid crystals: S. santhosh et al.},\ }\href@noop {} {\bibfield  {journal}
  {\bibinfo  {journal} {Journal of Statistical Physics}\ }\textbf {\bibinfo
  {volume} {180}},\ \bibinfo {pages} {699} (\bibinfo {year}
  {2020})}\BibitemShut {NoStop}%
\bibitem [{\citenamefont {Srivastava}\ \emph {et~al.}(2016)\citenamefont
  {Srivastava}, \citenamefont {Mishra},\ and\ \citenamefont
  {Marchetti}}]{srivastava2016negative}%
  \BibitemOpen
  \bibfield  {author} {\bibinfo {author} {\bibfnamefont {P.}~\bibnamefont
  {Srivastava}}, \bibinfo {author} {\bibfnamefont {P.}~\bibnamefont {Mishra}},\
  and\ \bibinfo {author} {\bibfnamefont {M.~C.}\ \bibnamefont {Marchetti}},\
  }\bibfield  {title} {\bibinfo {title} {Negative stiffness and modulated
  states in active nematics},\ }\href@noop {} {\bibfield  {journal} {\bibinfo
  {journal} {Soft matter}\ }\textbf {\bibinfo {volume} {12}},\ \bibinfo {pages}
  {8214} (\bibinfo {year} {2016})}\BibitemShut {NoStop}%
\bibitem [{\citenamefont {Krommydas}\ \emph {et~al.}(2026)\citenamefont
  {Krommydas}, \citenamefont {Gulati}, \citenamefont {Baskaran},\ and\
  \citenamefont {Marchetti}}]{krommydas2026controlling}%
  \BibitemOpen
  \bibfield  {author} {\bibinfo {author} {\bibfnamefont {D.}~\bibnamefont
  {Krommydas}}, \bibinfo {author} {\bibfnamefont {P.}~\bibnamefont {Gulati}},
  \bibinfo {author} {\bibfnamefont {A.}~\bibnamefont {Baskaran}},\ and\
  \bibinfo {author} {\bibfnamefont {M.~C.}\ \bibnamefont {Marchetti}},\
  }\bibfield  {title} {\bibinfo {title} {Controlling turbulent flows in
  compressible active nematics},\ }\href
  {https://doi.org/10.48550/arXiv.2607.24927} {\bibfield  {journal} {\bibinfo
  {journal} {arXiv}\ ,\ \bibinfo {eid} {2607.24927}} (\bibinfo {year}
  {2026})}\BibitemShut {NoStop}%
\bibitem [{\citenamefont {Marenduzzo}\ \emph {et~al.}(2007)\citenamefont
  {Marenduzzo}, \citenamefont {Orlandini}, \citenamefont {Cates},\ and\
  \citenamefont {Yeomans}}]{marenduzzo2007steady}%
  \BibitemOpen
  \bibfield  {author} {\bibinfo {author} {\bibfnamefont {D.}~\bibnamefont
  {Marenduzzo}}, \bibinfo {author} {\bibfnamefont {E.}~\bibnamefont
  {Orlandini}}, \bibinfo {author} {\bibfnamefont {M.~E.}\ \bibnamefont
  {Cates}},\ and\ \bibinfo {author} {\bibfnamefont {J.~M.}\ \bibnamefont
  {Yeomans}},\ }\bibfield  {title} {\bibinfo {title} {Steady-state hydrodynamic
  instabilities of active liquid crystals: Hybrid lattice boltzmann
  simulations.},\ }\href {https://doi.org/10.1103/PhysRevE.76.031921}
  {\bibfield  {journal} {\bibinfo  {journal} {Phys. Rev. E}\ }\textbf {\bibinfo
  {volume} {76}},\ \bibinfo {pages} {031921} (\bibinfo {year}
  {2007})}\BibitemShut {NoStop}%
\bibitem [{\citenamefont {Giomi}\ \emph {et~al.}(2014)\citenamefont {Giomi},
  \citenamefont {Bowick}, \citenamefont {Mishra}, \citenamefont {Sknepnek},\
  and\ \citenamefont {Marchetti}}]{giomi2014defect}%
  \BibitemOpen
  \bibfield  {author} {\bibinfo {author} {\bibfnamefont {L.}~\bibnamefont
  {Giomi}}, \bibinfo {author} {\bibfnamefont {M.~J.}\ \bibnamefont {Bowick}},
  \bibinfo {author} {\bibfnamefont {P.}~\bibnamefont {Mishra}}, \bibinfo
  {author} {\bibfnamefont {R.}~\bibnamefont {Sknepnek}},\ and\ \bibinfo
  {author} {\bibfnamefont {M.~C.}\ \bibnamefont {Marchetti}},\ }\bibfield
  {title} {\bibinfo {title} {Defect dynamics in active nematics},\ }\href
  {https://doi.org/10.1098/rsta.2013.0365} {\bibfield  {journal} {\bibinfo
  {journal} {Philos. Trans. R. Soc. A}\ }\textbf {\bibinfo {volume} {372}},\
  \bibinfo {pages} {20130365} (\bibinfo {year} {2014})}\BibitemShut {NoStop}%
\bibitem [{\citenamefont {Giomi}(2015)}]{giomi2015geometry}%
  \BibitemOpen
  \bibfield  {author} {\bibinfo {author} {\bibfnamefont {L.}~\bibnamefont
  {Giomi}},\ }\bibfield  {title} {\bibinfo {title} {Geometry and topology of
  turbulence in active nematics},\ }\href
  {https://doi.org/10.1103/PhysRevX.5.031003} {\bibfield  {journal} {\bibinfo
  {journal} {Phys. Rev. X}\ }\textbf {\bibinfo {volume} {5}},\ \bibinfo {pages}
  {031003} (\bibinfo {year} {2015})}\BibitemShut {NoStop}%
\bibitem [{\citenamefont {Ruske}\ and\ \citenamefont
  {Yeomans}(2022)}]{ruske2022activity}%
  \BibitemOpen
  \bibfield  {author} {\bibinfo {author} {\bibfnamefont {L.~J.}\ \bibnamefont
  {Ruske}}\ and\ \bibinfo {author} {\bibfnamefont {J.~M.}\ \bibnamefont
  {Yeomans}},\ }\bibfield  {title} {\bibinfo {title} {Activity gradients in
  two- and three-dimensional active nematics},\ }\href
  {https://doi.org/10.1039/D2SM00228K} {\bibfield  {journal} {\bibinfo
  {journal} {Soft Matter}\ }\textbf {\bibinfo {volume} {18}},\ \bibinfo {pages}
  {5654} (\bibinfo {year} {2022})}\BibitemShut {NoStop}%
\bibitem [{\citenamefont {Schnitzer}(1993)}]{schnitzer1993theory}%
  \BibitemOpen
  \bibfield  {author} {\bibinfo {author} {\bibfnamefont {M.~J.}\ \bibnamefont
  {Schnitzer}},\ }\bibfield  {title} {\bibinfo {title} {Theory of continuum
  random walks and application to chemotaxis},\ }\href
  {https://doi.org/10.1103/PhysRevE.48.2553} {\bibfield  {journal} {\bibinfo
  {journal} {Phys. Rev. E}\ }\textbf {\bibinfo {volume} {48}},\ \bibinfo
  {pages} {2553} (\bibinfo {year} {1993})}\BibitemShut {NoStop}%
\bibitem [{\citenamefont {Solon}\ \emph {et~al.}(2018)\citenamefont {Solon},
  \citenamefont {Stenhammar}, \citenamefont {Cates}, \citenamefont {Kafri},\
  and\ \citenamefont {Tailleur}}]{solon2018generalized}%
  \BibitemOpen
  \bibfield  {author} {\bibinfo {author} {\bibfnamefont {A.~P.}\ \bibnamefont
  {Solon}}, \bibinfo {author} {\bibfnamefont {J.}~\bibnamefont {Stenhammar}},
  \bibinfo {author} {\bibfnamefont {M.~E.}\ \bibnamefont {Cates}}, \bibinfo
  {author} {\bibfnamefont {Y.}~\bibnamefont {Kafri}},\ and\ \bibinfo {author}
  {\bibfnamefont {J.}~\bibnamefont {Tailleur}},\ }\bibfield  {title} {\bibinfo
  {title} {Generalized thermodynamics of phase equilibria in scalar active
  matter},\ }\href {https://doi.org/10.1103/PhysRevE.97.020602} {\bibfield
  {journal} {\bibinfo  {journal} {Phys. Rev. E}\ }\textbf {\bibinfo {volume}
  {97}},\ \bibinfo {pages} {020602} (\bibinfo {year} {2018})}\BibitemShut
  {NoStop}%
\bibitem [{\citenamefont {Fily}\ and\ \citenamefont
  {Marchetti}(2012)}]{fily2012athermal}%
  \BibitemOpen
  \bibfield  {author} {\bibinfo {author} {\bibfnamefont {Y.}~\bibnamefont
  {Fily}}\ and\ \bibinfo {author} {\bibfnamefont {M.~C.}\ \bibnamefont
  {Marchetti}},\ }\bibfield  {title} {\bibinfo {title} {Athermal phase
  separation of self-propelled particles with no alignment},\ }\href
  {https://doi.org/10.1103/PhysRevLett.108.235702} {\bibfield  {journal}
  {\bibinfo  {journal} {Phys. Rev. Lett.}\ }\textbf {\bibinfo {volume} {108}},\
  \bibinfo {pages} {235702} (\bibinfo {year} {2012})}\BibitemShut {NoStop}%
\bibitem [{\citenamefont {Marchetti}\ \emph {et~al.}(2016)\citenamefont
  {Marchetti}, \citenamefont {Fily}, \citenamefont {Henkes}, \citenamefont
  {Patch},\ and\ \citenamefont {Yllanes}}]{marchetti2016minimal}%
  \BibitemOpen
  \bibfield  {author} {\bibinfo {author} {\bibfnamefont {M.~C.}\ \bibnamefont
  {Marchetti}}, \bibinfo {author} {\bibfnamefont {Y.}~\bibnamefont {Fily}},
  \bibinfo {author} {\bibfnamefont {S.}~\bibnamefont {Henkes}}, \bibinfo
  {author} {\bibfnamefont {A.}~\bibnamefont {Patch}},\ and\ \bibinfo {author}
  {\bibfnamefont {D.}~\bibnamefont {Yllanes}},\ }\bibfield  {title} {\bibinfo
  {title} {Minimal model of active colloids highlights the role of mechanical
  interactions in controlling the emergent behavior of active matter},\ }\href
  {https://doi.org/10.1016/j.cocis.2016.01.003} {\bibfield  {journal} {\bibinfo
   {journal} {Curr. Opin. Colloid Interface Sci.}\ }\textbf {\bibinfo {volume}
  {21}},\ \bibinfo {pages} {34} (\bibinfo {year} {2016})}\BibitemShut {NoStop}%
\bibitem [{\citenamefont {Toner}\ and\ \citenamefont
  {Tu}(1995)}]{toner1995long}%
  \BibitemOpen
  \bibfield  {author} {\bibinfo {author} {\bibfnamefont {J.}~\bibnamefont
  {Toner}}\ and\ \bibinfo {author} {\bibfnamefont {Y.}~\bibnamefont {Tu}},\
  }\bibfield  {title} {\bibinfo {title} {Long-range order in a two-dimensional
  dynamical {XY} model: How birds fly together},\ }\href
  {https://doi.org/10.1103/PhysRevLett.75.4326} {\bibfield  {journal} {\bibinfo
   {journal} {Phys. Rev. Lett.}\ }\textbf {\bibinfo {volume} {75}},\ \bibinfo
  {pages} {4326} (\bibinfo {year} {1995})}\BibitemShut {NoStop}%
\bibitem [{\citenamefont {Cates}\ and\ \citenamefont
  {Tailleur}(2015)}]{cates2015motility}%
  \BibitemOpen
  \bibfield  {author} {\bibinfo {author} {\bibfnamefont {M.~E.}\ \bibnamefont
  {Cates}}\ and\ \bibinfo {author} {\bibfnamefont {J.}~\bibnamefont
  {Tailleur}},\ }\bibfield  {title} {\bibinfo {title} {Motility-induced phase
  separation},\ }\href
  {https://doi.org/10.1146/annurev-conmatphys-031214-014710} {\bibfield
  {journal} {\bibinfo  {journal} {Annu. Rev. Condens. Matter Phys.}\ }\textbf
  {\bibinfo {volume} {6}},\ \bibinfo {pages} {219} (\bibinfo {year}
  {2015})}\BibitemShut {NoStop}%
\bibitem [{\citenamefont {Maryshev}\ \emph {et~al.}(2019)\citenamefont
  {Maryshev}, \citenamefont {Goryachev}, \citenamefont {Marenduzzo},\ and\
  \citenamefont {Morozov}}]{maryshev2019dry}%
  \BibitemOpen
  \bibfield  {author} {\bibinfo {author} {\bibfnamefont {I.}~\bibnamefont
  {Maryshev}}, \bibinfo {author} {\bibfnamefont {A.~B.}\ \bibnamefont
  {Goryachev}}, \bibinfo {author} {\bibfnamefont {D.}~\bibnamefont
  {Marenduzzo}},\ and\ \bibinfo {author} {\bibfnamefont {A.}~\bibnamefont
  {Morozov}},\ }\bibfield  {title} {\bibinfo {title} {Dry active turbulence in
  a model for microtubule--motor mixtures},\ }\href
  {https://doi.org/10.1039/C9SM00558G} {\bibfield  {journal} {\bibinfo
  {journal} {Soft Matter}\ }\textbf {\bibinfo {volume} {15}},\ \bibinfo {pages}
  {6038} (\bibinfo {year} {2019})}\BibitemShut {NoStop}%
\bibitem [{\citenamefont {Chat{\'e}}\ and\ \citenamefont
  {Mahault}(2022)}]{chate2022dry}%
  \BibitemOpen
  \bibfield  {author} {\bibinfo {author} {\bibfnamefont {H.}~\bibnamefont
  {Chat{\'e}}}\ and\ \bibinfo {author} {\bibfnamefont {B.}~\bibnamefont
  {Mahault}},\ }\bibfield  {title} {\bibinfo {title} {Dry, aligning, dilute,
  active matter: A synthetic and self-contained overview},\ }in\ \href
  {https://doi.org/10.1093/oso/9780192858313.003.0001} {\emph {\bibinfo
  {booktitle} {Active Matter and Nonequilibrium Statistical Physics: Lecture
  Notes of the Les Houches Summer School, Volume 112}}},\ \bibinfo {editor}
  {edited by\ \bibinfo {editor} {\bibfnamefont {G.}~\bibnamefont {Gompper}},
  \bibinfo {editor} {\bibfnamefont {M.~C.}\ \bibnamefont {Marchetti}}, \bibinfo
  {editor} {\bibfnamefont {J.}~\bibnamefont {Tailleur}}, \bibinfo {editor}
  {\bibfnamefont {J.~M.}\ \bibnamefont {Yeomans}},\ and\ \bibinfo {editor}
  {\bibfnamefont {C.}~\bibnamefont {Salomon}}}\ (\bibinfo  {publisher} {Oxford
  University Press},\ \bibinfo {year} {2022})\ pp.\ \bibinfo {pages}
  {3--51}\BibitemShut {NoStop}%
\bibitem [{\citenamefont {Doostmohammadi}\ \emph {et~al.}(2016)\citenamefont
  {Doostmohammadi}, \citenamefont {Adamer}, \citenamefont {Thampi},\ and\
  \citenamefont {Yeomans}}]{doostmohammadi2016stabilization}%
  \BibitemOpen
  \bibfield  {author} {\bibinfo {author} {\bibfnamefont {A.}~\bibnamefont
  {Doostmohammadi}}, \bibinfo {author} {\bibfnamefont {M.~F.}\ \bibnamefont
  {Adamer}}, \bibinfo {author} {\bibfnamefont {S.~P.}\ \bibnamefont {Thampi}},\
  and\ \bibinfo {author} {\bibfnamefont {J.~M.}\ \bibnamefont {Yeomans}},\
  }\bibfield  {title} {\bibinfo {title} {Stabilization of active matter by
  flow-vortex lattices and defect ordering},\ }\href
  {https://doi.org/10.1038/ncomms10557} {\bibfield  {journal} {\bibinfo
  {journal} {Nat. Commun.}\ }\textbf {\bibinfo {volume} {7}},\ \bibinfo {pages}
  {10557} (\bibinfo {year} {2016})}\BibitemShut {NoStop}%
\bibitem [{\citenamefont {Bhattacharyya}\ and\ \citenamefont
  {Yeomans}(2023)}]{bhattacharyya2023phase}%
  \BibitemOpen
  \bibfield  {author} {\bibinfo {author} {\bibfnamefont {S.}~\bibnamefont
  {Bhattacharyya}}\ and\ \bibinfo {author} {\bibfnamefont {J.~M.}\ \bibnamefont
  {Yeomans}},\ }\bibfield  {title} {\bibinfo {title} {Phase separation driven
  by active flows},\ }\href@noop {} {\bibfield  {journal} {\bibinfo  {journal}
  {Physical Review Letters}\ }\textbf {\bibinfo {volume} {130}},\ \bibinfo
  {pages} {238201} (\bibinfo {year} {2023})}\BibitemShut {NoStop}%
\bibitem [{\citenamefont {Bhattacharyya}\ and\ \citenamefont
  {Yeomans}(2025)}]{bhattacharyya2025active}%
  \BibitemOpen
  \bibfield  {author} {\bibinfo {author} {\bibfnamefont {S.}~\bibnamefont
  {Bhattacharyya}}\ and\ \bibinfo {author} {\bibfnamefont {J.~M.}\ \bibnamefont
  {Yeomans}},\ }\bibfield  {title} {\bibinfo {title} {Active sorting to
  boundaries in active nematic--passive isotropic fluid mixtures},\ }\href
  {https://doi.org/10.1039/D5SM00636H} {\bibfield  {journal} {\bibinfo
  {journal} {Soft Matter}\ }\textbf {\bibinfo {volume} {21}},\ \bibinfo {pages}
  {7641} (\bibinfo {year} {2025})}\BibitemShut {NoStop}%
\bibitem [{\citenamefont {Prost}\ \emph {et~al.}(2015)\citenamefont {Prost},
  \citenamefont {J{\"u}licher},\ and\ \citenamefont
  {Joanny}}]{prost2015active}%
  \BibitemOpen
  \bibfield  {author} {\bibinfo {author} {\bibfnamefont {J.}~\bibnamefont
  {Prost}}, \bibinfo {author} {\bibfnamefont {F.}~\bibnamefont
  {J{\"u}licher}},\ and\ \bibinfo {author} {\bibfnamefont {J.-F.}\ \bibnamefont
  {Joanny}},\ }\bibfield  {title} {\bibinfo {title} {Active gel physics},\
  }\href {https://doi.org/10.1038/nphys3224} {\bibfield  {journal} {\bibinfo
  {journal} {Nat. Phys.}\ }\textbf {\bibinfo {volume} {11}},\ \bibinfo {pages}
  {111} (\bibinfo {year} {2015})}\BibitemShut {NoStop}%
\bibitem [{\citenamefont {Wioland}\ \emph
  {et~al.}(2013{\natexlab{b}})\citenamefont {Wioland}, \citenamefont
  {Woodhouse}, \citenamefont {Dunkel}, \citenamefont {Kessler},\ and\
  \citenamefont {Goldstein}}]{Wioland2013}%
  \BibitemOpen
  \bibfield  {author} {\bibinfo {author} {\bibfnamefont {H.}~\bibnamefont
  {Wioland}}, \bibinfo {author} {\bibfnamefont {F.~G.}\ \bibnamefont
  {Woodhouse}}, \bibinfo {author} {\bibfnamefont {J.}~\bibnamefont {Dunkel}},
  \bibinfo {author} {\bibfnamefont {J.~O.}\ \bibnamefont {Kessler}},\ and\
  \bibinfo {author} {\bibfnamefont {R.~E.}\ \bibnamefont {Goldstein}},\
  }\bibfield  {title} {\bibinfo {title} {Confinement stabilizes a bacterial
  suspension into a spiral vortex},\ }\href
  {https://doi.org/10.1103/PhysRevLett.110.268102} {\bibfield  {journal}
  {\bibinfo  {journal} {Phys. Rev. Lett.}\ }\textbf {\bibinfo {volume} {110}},\
  \bibinfo {pages} {268102} (\bibinfo {year} {2013}{\natexlab{b}})}\BibitemShut
  {NoStop}%
\bibitem [{\citenamefont {Wu}\ \emph {et~al.}(2017{\natexlab{b}})\citenamefont
  {Wu}, \citenamefont {Hishamunda}, \citenamefont {Chen}, \citenamefont
  {DeCamp}, \citenamefont {Chang}, \citenamefont {Fern{\'a}ndez-Nieves},
  \citenamefont {Fraden},\ and\ \citenamefont {Dogic}}]{Wu2017}%
  \BibitemOpen
  \bibfield  {author} {\bibinfo {author} {\bibfnamefont {K.-T.}\ \bibnamefont
  {Wu}}, \bibinfo {author} {\bibfnamefont {J.~B.}\ \bibnamefont {Hishamunda}},
  \bibinfo {author} {\bibfnamefont {D.~T.~N.}\ \bibnamefont {Chen}}, \bibinfo
  {author} {\bibfnamefont {S.~J.}\ \bibnamefont {DeCamp}}, \bibinfo {author}
  {\bibfnamefont {Y.-W.}\ \bibnamefont {Chang}}, \bibinfo {author}
  {\bibfnamefont {A.}~\bibnamefont {Fern{\'a}ndez-Nieves}}, \bibinfo {author}
  {\bibfnamefont {S.}~\bibnamefont {Fraden}},\ and\ \bibinfo {author}
  {\bibfnamefont {Z.}~\bibnamefont {Dogic}},\ }\bibfield  {title} {\bibinfo
  {title} {Transition from turbulent to coherent flows in confined
  three-dimensional active fluids},\ }\href
  {https://doi.org/10.1126/science.aal1979} {\bibfield  {journal} {\bibinfo
  {journal} {Science}\ }\textbf {\bibinfo {volume} {355}},\ \bibinfo {pages}
  {eaal1979} (\bibinfo {year} {2017}{\natexlab{b}})}\BibitemShut {NoStop}%
\bibitem [{\citenamefont {Nishiguchi}\ \emph {et~al.}(2018)\citenamefont
  {Nishiguchi}, \citenamefont {Aranson}, \citenamefont {Snezhko},\ and\
  \citenamefont {Sokolov}}]{Nishiguchi2018}%
  \BibitemOpen
  \bibfield  {author} {\bibinfo {author} {\bibfnamefont {D.}~\bibnamefont
  {Nishiguchi}}, \bibinfo {author} {\bibfnamefont {I.~S.}\ \bibnamefont
  {Aranson}}, \bibinfo {author} {\bibfnamefont {A.}~\bibnamefont {Snezhko}},\
  and\ \bibinfo {author} {\bibfnamefont {A.}~\bibnamefont {Sokolov}},\
  }\bibfield  {title} {\bibinfo {title} {Engineering bacterial vortex lattice
  via direct laser lithography},\ }\href
  {https://doi.org/10.1038/s41467-018-06842-6} {\bibfield  {journal} {\bibinfo
  {journal} {Nat. Commun.}\ }\textbf {\bibinfo {volume} {9}},\ \bibinfo {pages}
  {4486} (\bibinfo {year} {2018})}\BibitemShut {NoStop}%
\bibitem [{\citenamefont {Hardo{\"u}in}\ \emph {et~al.}(2019)\citenamefont
  {Hardo{\"u}in}, \citenamefont {Hughes}, \citenamefont {Doostmohammadi},
  \citenamefont {Laurent}, \citenamefont {Lopez-Leon}, \citenamefont {Yeomans},
  \citenamefont {Ign{\'e}s-Mullol},\ and\ \citenamefont
  {Sagu{\'e}s}}]{Hardouin2019}%
  \BibitemOpen
  \bibfield  {author} {\bibinfo {author} {\bibfnamefont {J.}~\bibnamefont
  {Hardo{\"u}in}}, \bibinfo {author} {\bibfnamefont {R.}~\bibnamefont
  {Hughes}}, \bibinfo {author} {\bibfnamefont {A.}~\bibnamefont
  {Doostmohammadi}}, \bibinfo {author} {\bibfnamefont {J.}~\bibnamefont
  {Laurent}}, \bibinfo {author} {\bibfnamefont {T.}~\bibnamefont {Lopez-Leon}},
  \bibinfo {author} {\bibfnamefont {J.~M.}\ \bibnamefont {Yeomans}}, \bibinfo
  {author} {\bibfnamefont {J.}~\bibnamefont {Ign{\'e}s-Mullol}},\ and\ \bibinfo
  {author} {\bibfnamefont {F.}~\bibnamefont {Sagu{\'e}s}},\ }\bibfield  {title}
  {\bibinfo {title} {Reconfigurable flows and defect landscape of confined
  active nematics},\ }\href {https://doi.org/10.1038/s42005-019-0221-x}
  {\bibfield  {journal} {\bibinfo  {journal} {Commun. Phys.}\ }\textbf
  {\bibinfo {volume} {2}},\ \bibinfo {pages} {121} (\bibinfo {year}
  {2019})}\BibitemShut {NoStop}%
\bibitem [{\citenamefont {Opathalage}\ \emph {et~al.}(2019)\citenamefont
  {Opathalage}, \citenamefont {Norton}, \citenamefont {Juniper}, \citenamefont
  {Langeslay}, \citenamefont {Aghvami}, \citenamefont {Fraden},\ and\
  \citenamefont {Dogic}}]{Opathalage2019}%
  \BibitemOpen
  \bibfield  {author} {\bibinfo {author} {\bibfnamefont {A.}~\bibnamefont
  {Opathalage}}, \bibinfo {author} {\bibfnamefont {M.~M.}\ \bibnamefont
  {Norton}}, \bibinfo {author} {\bibfnamefont {M.~P.~N.}\ \bibnamefont
  {Juniper}}, \bibinfo {author} {\bibfnamefont {B.}~\bibnamefont {Langeslay}},
  \bibinfo {author} {\bibfnamefont {S.~A.}\ \bibnamefont {Aghvami}}, \bibinfo
  {author} {\bibfnamefont {S.}~\bibnamefont {Fraden}},\ and\ \bibinfo {author}
  {\bibfnamefont {Z.}~\bibnamefont {Dogic}},\ }\bibfield  {title} {\bibinfo
  {title} {Self-organized dynamics and the transition to turbulence of confined
  active nematics},\ }\href {https://doi.org/10.1073/pnas.1816733116}
  {\bibfield  {journal} {\bibinfo  {journal} {Proc. Natl. Acad. Sci. U.S.A.}\
  }\textbf {\bibinfo {volume} {116}},\ \bibinfo {pages} {4788} (\bibinfo {year}
  {2019})}\BibitemShut {NoStop}%
\bibitem [{\citenamefont {Chandrakar}\ \emph {et~al.}(2020)\citenamefont
  {Chandrakar}, \citenamefont {Varghese}, \citenamefont {Aghvami},
  \citenamefont {Baskaran}, \citenamefont {Dogic},\ and\ \citenamefont
  {Duclos}}]{Chandrakar2020}%
  \BibitemOpen
  \bibfield  {author} {\bibinfo {author} {\bibfnamefont {P.}~\bibnamefont
  {Chandrakar}}, \bibinfo {author} {\bibfnamefont {M.}~\bibnamefont
  {Varghese}}, \bibinfo {author} {\bibfnamefont {S.~A.}\ \bibnamefont
  {Aghvami}}, \bibinfo {author} {\bibfnamefont {A.}~\bibnamefont {Baskaran}},
  \bibinfo {author} {\bibfnamefont {Z.}~\bibnamefont {Dogic}},\ and\ \bibinfo
  {author} {\bibfnamefont {G.}~\bibnamefont {Duclos}},\ }\bibfield  {title}
  {\bibinfo {title} {Confinement controls the bend instability of
  three-dimensional active liquid crystals},\ }\href
  {https://doi.org/10.1103/PhysRevLett.125.257801} {\bibfield  {journal}
  {\bibinfo  {journal} {Phys. Rev. Lett.}\ }\textbf {\bibinfo {volume} {125}},\
  \bibinfo {pages} {257801} (\bibinfo {year} {2020})}\BibitemShut {NoStop}%
\bibitem [{\citenamefont {Buttinoni}\ \emph {et~al.}(2013)\citenamefont
  {Buttinoni}, \citenamefont {Bialk{\'e}}, \citenamefont {K{\"u}mmel},
  \citenamefont {L{\"o}wen}, \citenamefont {Bechinger},\ and\ \citenamefont
  {Speck}}]{Buttinoni2013}%
  \BibitemOpen
  \bibfield  {author} {\bibinfo {author} {\bibfnamefont {I.}~\bibnamefont
  {Buttinoni}}, \bibinfo {author} {\bibfnamefont {J.}~\bibnamefont
  {Bialk{\'e}}}, \bibinfo {author} {\bibfnamefont {F.}~\bibnamefont
  {K{\"u}mmel}}, \bibinfo {author} {\bibfnamefont {H.}~\bibnamefont
  {L{\"o}wen}}, \bibinfo {author} {\bibfnamefont {C.}~\bibnamefont
  {Bechinger}},\ and\ \bibinfo {author} {\bibfnamefont {T.}~\bibnamefont
  {Speck}},\ }\bibfield  {title} {\bibinfo {title} {Dynamical clustering and
  phase separation in suspensions of self-propelled colloidal particles},\
  }\href {https://doi.org/10.1103/PhysRevLett.110.238301} {\bibfield  {journal}
  {\bibinfo  {journal} {Phys. Rev. Lett.}\ }\textbf {\bibinfo {volume} {110}},\
  \bibinfo {pages} {238301} (\bibinfo {year} {2013})}\BibitemShut {NoStop}%
\bibitem [{\citenamefont {Maggi}\ \emph {et~al.}(2018)\citenamefont {Maggi},
  \citenamefont {Angelani}, \citenamefont {Frangipane},\ and\ \citenamefont
  {Di~Leonardo}}]{maggi2018currents}%
  \BibitemOpen
  \bibfield  {author} {\bibinfo {author} {\bibfnamefont {C.}~\bibnamefont
  {Maggi}}, \bibinfo {author} {\bibfnamefont {L.}~\bibnamefont {Angelani}},
  \bibinfo {author} {\bibfnamefont {G.}~\bibnamefont {Frangipane}},\ and\
  \bibinfo {author} {\bibfnamefont {R.}~\bibnamefont {Di~Leonardo}},\
  }\bibfield  {title} {\bibinfo {title} {Currents and flux-inversion in
  photokinetic active particles},\ }\href {https://doi.org/10.1039/C8SM00788H}
  {\bibfield  {journal} {\bibinfo  {journal} {Soft Matter}\ }\textbf {\bibinfo
  {volume} {14}},\ \bibinfo {pages} {4958} (\bibinfo {year}
  {2018})}\BibitemShut {NoStop}%
\bibitem [{\citenamefont {Zhao}\ \emph {et~al.}(2023)\citenamefont {Zhao},
  \citenamefont {Ko{\v{s}}mrlj},\ and\ \citenamefont
  {Datta}}]{zhao2023chemotactic}%
  \BibitemOpen
  \bibfield  {author} {\bibinfo {author} {\bibfnamefont {H.}~\bibnamefont
  {Zhao}}, \bibinfo {author} {\bibfnamefont {A.}~\bibnamefont
  {Ko{\v{s}}mrlj}},\ and\ \bibinfo {author} {\bibfnamefont {S.~S.}\
  \bibnamefont {Datta}},\ }\bibfield  {title} {\bibinfo {title} {Chemotactic
  motility-induced phase separation},\ }\href
  {https://doi.org/10.1103/PhysRevLett.131.118301} {\bibfield  {journal}
  {\bibinfo  {journal} {Phys. Rev. Lett.}\ }\textbf {\bibinfo {volume} {131}},\
  \bibinfo {pages} {118301} (\bibinfo {year} {2023})}\BibitemShut {NoStop}%
\bibitem [{\citenamefont {Bai}\ \emph {et~al.}(2021)\citenamefont {Bai},
  \citenamefont {He}, \citenamefont {Chu}, \citenamefont {Long}, \citenamefont
  {Li},\ and\ \citenamefont {Fu}}]{bai2021spatial}%
  \BibitemOpen
  \bibfield  {author} {\bibinfo {author} {\bibfnamefont {Y.}~\bibnamefont
  {Bai}}, \bibinfo {author} {\bibfnamefont {C.}~\bibnamefont {He}}, \bibinfo
  {author} {\bibfnamefont {P.}~\bibnamefont {Chu}}, \bibinfo {author}
  {\bibfnamefont {J.}~\bibnamefont {Long}}, \bibinfo {author} {\bibfnamefont
  {X.}~\bibnamefont {Li}},\ and\ \bibinfo {author} {\bibfnamefont
  {X.}~\bibnamefont {Fu}},\ }\bibfield  {title} {\bibinfo {title} {Spatial
  modulation of individual behaviors enables an ordered structure of diverse
  phenotypes during bacterial group migration},\ }\href
  {https://doi.org/10.7554/eLife.67316} {\bibfield  {journal} {\bibinfo
  {journal} {eLife}\ }\textbf {\bibinfo {volume} {10}},\ \bibinfo {pages}
  {e67316} (\bibinfo {year} {2021})}\BibitemShut {NoStop}%
\bibitem [{\citenamefont {Tailleur}\ and\ \citenamefont
  {Cates}(2008)}]{TailleurCates2008}%
  \BibitemOpen
  \bibfield  {author} {\bibinfo {author} {\bibfnamefont {J.}~\bibnamefont
  {Tailleur}}\ and\ \bibinfo {author} {\bibfnamefont {M.~E.}\ \bibnamefont
  {Cates}},\ }\bibfield  {title} {\bibinfo {title} {Statistical mechanics of
  interacting run-and-tumble bacteria},\ }\href
  {https://doi.org/10.1103/PhysRevLett.100.218103} {\bibfield  {journal}
  {\bibinfo  {journal} {Phys. Rev. Lett.}\ }\textbf {\bibinfo {volume} {100}},\
  \bibinfo {pages} {218103} (\bibinfo {year} {2008})}\BibitemShut {NoStop}%
\bibitem [{\citenamefont {Stenhammar}\ \emph {et~al.}(2016)\citenamefont
  {Stenhammar}, \citenamefont {Wittkowski}, \citenamefont {Marenduzzo},\ and\
  \citenamefont {Cates}}]{Stenhammar2016}%
  \BibitemOpen
  \bibfield  {author} {\bibinfo {author} {\bibfnamefont {J.}~\bibnamefont
  {Stenhammar}}, \bibinfo {author} {\bibfnamefont {R.}~\bibnamefont
  {Wittkowski}}, \bibinfo {author} {\bibfnamefont {D.}~\bibnamefont
  {Marenduzzo}},\ and\ \bibinfo {author} {\bibfnamefont {M.~E.}\ \bibnamefont
  {Cates}},\ }\bibfield  {title} {\bibinfo {title} {Light-induced self-assembly
  of active rectification devices},\ }\href
  {https://doi.org/10.1126/sciadv.1501850} {\bibfield  {journal} {\bibinfo
  {journal} {Sci. Adv.}\ }\textbf {\bibinfo {volume} {2}},\ \bibinfo {pages}
  {e1501850} (\bibinfo {year} {2016})}\BibitemShut {NoStop}%
\bibitem [{\citenamefont {Tayar}\ \emph {et~al.}(2022)\citenamefont {Tayar},
  \citenamefont {Lemma},\ and\ \citenamefont {Dogic}}]{tayar2022assembling}%
  \BibitemOpen
  \bibfield  {author} {\bibinfo {author} {\bibfnamefont {A.~M.}\ \bibnamefont
  {Tayar}}, \bibinfo {author} {\bibfnamefont {L.~M.}\ \bibnamefont {Lemma}},\
  and\ \bibinfo {author} {\bibfnamefont {Z.}~\bibnamefont {Dogic}},\ }\bibfield
   {title} {\bibinfo {title} {Assembling microtubule-based active matter},\
  }\href {https://doi.org/10.1007/978-1-0716-1983-4_10} {\bibfield  {journal}
  {\bibinfo  {journal} {Methods Mol. Biol.}\ }\textbf {\bibinfo {volume}
  {2430}},\ \bibinfo {pages} {151} (\bibinfo {year} {2022})}\BibitemShut
  {NoStop}%
\bibitem [{\citenamefont {Lau}\ \emph {et~al.}(2009)\citenamefont {Lau},
  \citenamefont {Prasad},\ and\ \citenamefont {Dogic}}]{lau2009condensation}%
  \BibitemOpen
  \bibfield  {author} {\bibinfo {author} {\bibfnamefont {A.~W.~C.}\
  \bibnamefont {Lau}}, \bibinfo {author} {\bibfnamefont {A.}~\bibnamefont
  {Prasad}},\ and\ \bibinfo {author} {\bibfnamefont {Z.}~\bibnamefont
  {Dogic}},\ }\bibfield  {title} {\bibinfo {title} {Condensation of isolated
  semi-flexible filaments driven by depletion interactions},\ }\href
  {https://doi.org/10.1209/0295-5075/87/48006} {\bibfield  {journal} {\bibinfo
  {journal} {EPL}\ }\textbf {\bibinfo {volume} {87}},\ \bibinfo {pages} {48006}
  (\bibinfo {year} {2009})}\BibitemShut {NoStop}%
\end{thebibliography}%
\end{document}